\documentclass[AMA,Times1COL]{WileyNJDv5}

\articletype{}%
\copyyear{2024}
\startpage{1}

\begin{document}

\title{A Social Outcomes and Priorities centered (SOP) Framework for AI policy}

\author[1,2]{Mohak Shah}

\authormark{Shah}% \textsc{et al.}}
\titlemark{A Social Outcomes and Priorities centered (SOP) Framework for AI policy}

\address[1]{\orgname{Praescivi Advisors}, \orgaddress{\state{CA}, \country{USA}}}
\address[2]{\orgname{University of Illinois at Chicago}, \orgaddress{\state{IL}, \country{USA}}}

\corres{Corresponding author Mohak Shah. \email{mohak@mohakshah.com}}

\abstract[Abstract]{Rapid developments in AI and its adoption across various domains have necessitated a need to build robust guardrails and risk containment plans while ensuring equitable benefits for the betterment of society. The current technology-centered approach has resulted in a  fragmented, reactive, and ineffective policy apparatus. This paper highlights the immediate and urgent need to pivot to a \textit{society-centered approach} to develop comprehensive, coherent, forward-looking AI policy. To this end, we present a \textit{Social Outcomes and Priorities centered (SOP) framework for AI policy} along with proposals on implementation of its various components. While the SOP framework is presented from a US-centric view, the takeaways are general and applicable globally.}

\keywords{AI, AI Policy, AI Regulation, AI and Society, Generative AI}

\maketitle

\renewcommand\thefootnote{}
\footnotetext{\textbf{Abbreviations:} AI, Artificial Intelligence; GenAI, Generative AI; LLM, Large-language model; ML, Machine Learning; NIST, National Institute of Standards and Technology (US Dept. of Commerce); CFPB, Consumer Financial Protection Bureau; NHTSA, National Highway Traffic Safety Administration; DoD, Department of Defense; DARPA, Defense Advanced Research Projects Agency; AGI, Artificial General Intelligence; ASI, Artificial Super Intelligence; OECD, Organisation for Economic Co-operation and Development; EU, European Union; NSF, National Science Foundation; SEC, U.S. Securities and Exchange Commission; FTC, Federal Trade Commission; FDA, U.S. Food and Drug Administration; DoT, U.S. Department of Transportation; ADAS, Advanced Driver Assistance Systems; IoT, Internet of Things.}

\renewcommand\thefootnote{\fnsymbol{footnote}}
\setcounter{footnote}{1}

\section{Introduction}\label{sec-intro}

Developments in the field of AI have been rapid and continue to gather momentum. The application of AI spans a variety of domains in both public and private sectors, and continue to accelerate. Further, recent developments in AI such as Generative AI have resulted in a fundamental evolution of the types of AI systems -- systems that are not just stochastic but also manifest themselves in ways that exhibit previously unknown characteristics and behaviors (e.g., hallucinations~\cite{xu2024,banerjee-2024}). Approaches that rely on studying these systems as natural systems are bound to be very deficient in understanding their behavior, let alone quantifying them, since both the uses and coverage of these systems are neither limited nor well-understood. To add to that, the underlying system design isn't static and continues to evolve. We are sorely lacking a good understanding on the workings as well as proper, principled, tractable evaluation and validation of these AI systems.

Unlike even a couple of decades ago, this rapid evolution is accompanied by parallel developments in other relevant areas such as computing, semiconductors, large scale data availability, rapid adoption and integration capabilities, and relatively much shorter time to market -– i.e., reach to both enterprise and retail consumers. There are market competitive forces further pushing the AI technology makers (interestingly both in the industry and academia) to release products and services powered by these technologies at an increasing pace. Hence, the cumulative effects of these developments, market forces, and the competitive pressures are posing a unique set of challenges -- quite unlike what we have seen with critical technologies in the past where either the knowledge, ability to develop, or access (to the technology itself, or its building blocks) was typically restricted. 

Consequently, maturity of AI models and systems notwithstanding, their adoption has picked up significant pace~\cite{NBERw32966-2024} (the estimates in~\cite{NBERw32966-2024} are likely overly optimistic for GenAI adoption but the overall intent and industry efforts certainly continue to push for further penetration of technology, and other AI methods and techniques are already deeply entrenched in many areas). As adoption grows so do the challenges. These challenges are both specific and potentially systemic. More importantly, the debate on the effects and impact of AI on humanity tends to largely be held in abstract. On one end, proponents maintain AI as a type of silver bullet solution to everything that ails us, while the arguments on the other end maintain that AI poses existential risk in \textit{mostly overt ways}. It can be argued that AI poses systemic risks in various ways that can have significant deleterious implications for our society -- in fact they are manifesting already. However, most of the AI risk debate is around some hypothetical future scenario when AI systems will \textit{take over} and make decisions bypassing human control (an advanced AGI scenario which we are far from and neither have a realistic timeline nor understanding of) that can bring about catastrophic consequences on humanity -- scenarios such as nuclear warfare and bio-weapon risks.\footnote{These and other catastrophic risks are very real in today's world but currently contingent on human actions not a consequence of AI's automated-decision-making risk.} We humbly contend that these two extreme projections on the spectrum of AI's opportunities and risks not just misrepresent the stakes but distract us collectively from both much needed urgent actions and an informed lasting policy framework -- one that can address our present and immediate needs \textit{and} realistically and pragmatically scale with the growth of AI and other developments.

Neither the hyperbole around the all-encompassing benefits of AI, nor the doomsday scenarios of AGI taking over human decision making seem imminent and realistic enough currently so as to be actionable upon, but the risks and challenges arising from various relevant applications of AI are already here, and presently. Even when there are relatively more calibrated efforts in outlining the risks from autonomous AI systems~\cite{lumenova2024,bengio2024}, these are typically futuristic scenarios for which the current state and sophistication of AI doesn't suffice. And this is an important point. Both -- the intentional hype and/or overestimation of AI's current capability and the haste in putting immature, unreliable and untested AI capabilities in production to automate various critical tasks -- are themselves dangerous. They not just expand the risk vectors but also open up possibilities for unintended system failures leading to catastrophic consequences.

Among the present challenges and imminent risks from AI potentially in conjunction with other concurrent technological developments that confront us are:
\begin{itemize}
    \item Introduction of risks to products and services. Both due to a lack of proper testing framework and lack of incentive or requirement to do so, AI products and capabilities often lack robust testing, evaluation, validation and verification (V\&V). This results in a lack of rigorous vetting of the products entering the market and decision-making workflows. Among the most well-known recent examples of such risks are the hallucination and deepfake issue with GenAI products~\cite{katz2024, huang2023,kristof2024}. However, there are several additional examples leading to safety- and privacy-concerns among others~\cite{toulas2024, world-2023,swan2023,lecher2024}.
    \item Security vulnerability. Rapid introduction of half-baked products in business critical applications have opened up new fronts on data and security risks, and cyber-security lines of attacks. These risks are not just in terms of adversarial attacks and system compromise but also data compromise resulting from the manner in which the AI technologies are modeled and deployed~\cite{carlini2021,carlini2023,cohen2024,toulas2024,fu-2024}. 
    \item Impact on society and democracy. From potential for social engineering, election interference, mis- and dis-information, (misleading) partial information, to sowing division are only some of the risks that we need to contend with because of the combination of AI, mobile, and data technologies introduced and adopted across the world~\cite{pfau2024,coleman2024}.
    \item National security implications on important areas including IP, business competitiveness, cyber-security, and critical infrastructure~\cite{goodin2024,Heath2023,Bastian2023}.
    \item Lack of our collective capability to sufficiently understand and foresee, let alone address, the impact of AI-powered products and services be it on labor and workforce, social equity and equality, markets and consumers, or corporate responsibility~\cite{dubal2023,koebler2024,acemoglu2024}.
    \item Ethics and fairness challenges and concerns. These range all the way from data utilization, governance, and IP violation of human generated content to responsible use and societal implications around bias, discrimination, potential for social engineering, intended and unintended deleterious consequences~\cite{haim2024,walsh2023}.
    \item Risks and concerns emanating from relatively widespread open-source availability of novel AI technological developments (both in the US and globally)~\cite{seger-2023,harris-2024,owen-jackson-2024,ntia-2024,doerrfeld-2024}.
    \item Challenges from \textit{AI in conjunction with other technology areas} including quantum computing, blockchain, robotics, and Internet of Things (IoT) that may currently be escaping our focus and foresight~\cite{robey-2024,world-2023}.
    \item Less understood systemic and systematic effects of AI-driven products on existing functions and areas such as education, healthcare, psychology, and social harmony~\cite{striped2024,morrison-2023,robertson2024}.
\end{itemize}

Various policy efforts across the globe at both industry and government levels have been and continue to be proposed. While most of these efforts are mainly \textit{guidance} and lack teeth, the others are frameworks that focus on the specific sub-areas and/or technical aspects. All of them further are decoupled from the intended outcomes and lack clear definition and metrics of success. Consequently, we continue to discuss, hype, and claim advances on the AI policy front but have little to show for it in terms of intended social and societal results. Even when the proposals discuss the outlines of the expected outcomes or benefits to society, they are grossly missing in specifics and roadmap to achieve desired outcomes. Different policy elements and areas continue to be proposed independently without necessarily informing or coordinating with each other. Furthermore, we lack a core strategy to manage and reconcile them. Consequently, even though there is a lot of activity and efforts around building robust AI policies, standards, and guardrails, they are bound to miss in most cases since they are neither anchored in the expected outcomes nor systematically managed within a clear framework. 

Our current policy efforts fall under two broad categories. The first set of efforts focuses on placing very high-level, rather vague guidance that AI systems should benefit society and do no harm. While this is a worthy vision, it doesn't give any clarity or specifics on how to achieve that goal -- what outcomes and effects are desirable? What are the ones that society can adapt to and accommodate? Which outcomes are deleterious? Which are outright dangerous and risky, to be avoided at all costs?. There is also an inherent subjectivity involved in such specifics based on the application, domains, user-groups, sensitivity of the technologies involved, sensitivity and the level of risk involved, and short-term and accumulated risk scenarios. They do not clarify what type of overall policy approach can advance our objective of maximizing AI's benefits while minimizing the risks, and how it would be achieved. That is, we are missing a consensus social prioritization of the intended outcomes and an associated policy framework along with regulatory and standardization mechanism. Naturally, in the absence of any anchor or specificity, combined with the vagueness of guidance, these sets of policy suggestions have been rendered inactionable. As a consequence, most of the focus and energy of the AI policy discourse has been usurped by the second set of efforts - those focused just on the technical aspects of AI, resulting in a solely technology-centered approach. However, there is little on which this approach is anchored in terms of its ability to achieve the intended outcomes. Unfortunately, \textit{society} is mostly missing from the work on \textit{AI policy for society}.

What we need is a comprehensive policy framework guided by social priorities and associated outcomes. This should account for various aspects of AI development and deployment life-cycle, AI's implications on different application areas and domains, resultant effects and risks on various realms of society, implications on national and corporate security, accountability and risk mitigation from AI-driven offerings, and guardrails against unwanted and unintended near- and long-term consequences, all anchored in outcomes that support the societal priorities including democracy, fairness, equity and equality, information provenance, privacy, and security. That is, the \textit{intended outcomes} should inform the policy priorities and the use of AI in various areas. This would not just help us optimally secure against the risk of AI but will also help maximize the use and adoption of AI for social benefits. Moreover, the innovation in the field can be guided productively and efficiently against this backdrop of social priorities. 

As a simple example, hallucinations in LLM's is a well recognized problem and likely impossible to be completely solved in the way GenAI algorithms are designed currently~\cite{xu2024,banerjee-2024}. Hallucinations can pose a very high risk in various areas such as defense or safety-critical application and should be managed both in technical and utilization sense. However, there are areas where they may not pose a critical risk –- areas such as creative art (assuming concerns around IP and copyrighted materials are resolved). Hence, policy framework should be flexible on where and under what circumstances is AI adoption helpful and permissible, and under what conditions they should have guardrails or can be outright impermissible. Also, various application and domains would demand different levels of stringency of such guardrails but this can be achieved only if there is a clear understanding and agreement on what types of outcomes are desired and allowed. Similarly, AI recommender systems can have fewer guardrails when applied to shopping recommendation but significantly stringent ones for critical areas such as content recommendation on social media to minimize the unintended and undesired consequences. 

There is an additional dimension on permissible uses of AI despite the guardrails and an associated accountability and enforcement mechanism - that of adversarial or unethical actors. For instance, there are various proposals around dealing with deepfakes relying mostly on the developers adhering to the established protocols around using AI in a principled manner. This, of course, doesn't protect us from nefarious or unethical uses for which the legal framework remains unprepared. An example would be the role of deepfakes in degradation of women as detailed in~\cite{tiku2024,kristof2024}. Not only does this exemplify the unwanted use of AI but also the limits of our societal safeguards in extending protection against such uses. It would be very difficult to address such issues if the view on AI's use is entirely technology-centered and decoupled from broader social and societal priorities. The social implications and costs are immense and AI in such cases is just an enabling technology that accelerates the technical path for these unwanted efforts (see, for instance,~\cite{striped2024} for various dangers to the youth from social media that need our immediate and continued attention). A framework that reconciles our social priorities with the associated guardrails against AI's use as well as the accountability-, legal- and enforcement-apparatus is critically important.  

This paper proposes a Social-Outcomes and Priorities based (SOP) framework for AI Policy. \textit{The SOP framework re-frames the discussion of AI policy in a society-centered approach, a departure from a technology-centered approach that has been adopted so far.} The paper outlines the core components of the SOP framework along with suggestions on their implementation, recommendations on how the framework can leverage the existing policy, regulatory, and legal apparatus and further advises on how these can be enhanced, strengthened and updated in the context of an evolving AI landscape. 

The rest of the paper is organized as follows: Sec.~\ref{sec:current-ai-policy} provides background on the main arguments currently driving the policy discussion resulting in fragmented technology-centered efforts, and their limitations. Sec.~\ref{sec:constituents-of-ai-policy} sets up the discussion on the main constituents that an effective policy framework should cover. Sec.~\ref{sec:sop-framework} then introduces the main proposal -- a social-outcomes and priorities centered (SOP) framework for AI policy. It also details the main functional components of this framework along with proposals on their implementation. Even though the implementation proposals for various functions are presented from a US-centric viewpoint, the core framework is applicable globally. Sec.~\ref{sec:remarks} then details how an SOP framework can be beneficial and contextualizes it with the application. We finally conclude with a call for action in Sec.~\ref{sec:conclusion} highlighting both the urgency and importance of implementing such an effective policy framework.

\section{Current AI policy approaches}\label{sec:current-ai-policy}

The pace of developments in AI and other concurrent technology in conjunction with rapid adoption efforts have on the one end generated significant enthusiasm about its promise, while on the other hand has raised alarms around its risks. This trend has significantly picked up with the introduction of GenAI capabilities resulting in a host of novel challenges. However, while looking at implications of AI on society -- both in terms of promise and risks -- this is also an opportune and urgent time to address the broad AI space  that is already powering a range of applications in our daily lives. There is certainly a vigorous and global policy discourse around AI with various stakeholders taking positions ranging from the need to regulate certain developments (e.g.,~\cite{Schatz2021,cpra2024,Berman2024,Clarke2023}) in AI to the ones that oppose any AI regulations altogether(e.g.,~\cite{sharma-2023}). Further, most of the suggestions on building guardrails around AI and ensuring the right results are all centered on treating the policy effort \textit{solely} as a technological challenge. The central argument we make in this paper is that a technology-centric approach is insufficient and inappropriate when it comes to ascertain desirable social results and the stated policy objectives in most cases. Moreover, the current approach is resulting in an inadequate and ineffective policy apparatus. Before we highlight the shortcomings of this technology-centric approach, let us look at the main abstractions of the arguments that have played a key role in guiding the current policy approach, and the associated impact:

\begin{enumerate}
    \item \textit{Deep technical understanding as the sole basis for AI policy:} This argument goes along the lines that only deep technical understanding of AI can be the sole basis for forming any AI policy framework - in effect being the necessary and sufficient condition. This isn't just a flawed position but also results in most of the impacted stakeholders being left out from the debate  having no real say on the matter. While the AI SME's may have deep technical understanding of AI, they lack the rest of the contextual understanding and potentially the needed objectivity to propose informed policy positions for better societal outcomes.
    \item \textit{Regulating AI would be an Innovation killer:} There is a camp that opposes any type of regulatory efforts around AI~\cite{sharma-2023}. The argument being that any effort to build policies or regulations risks killing innovation. Passionate positions have been made from both side of this argument. However, we need a more nuanced discussion on the subject and ensure that meaningful innovation is encouraged while minimizing current and future risks. An entirely regulation-free environment evidently poses a significant risk to society and can have systemic deleterious effects.
    \item \textit{Self-regulation is the only way forward:} This argument line states that the industry should be allowed to set standards and self-regulate so that they can invest in building AI guardrails but without any kind of oversight or public-sector intervention. This is guaranteed to fail and has never in the past worked for any technology or industry mostly since without any input from public directly or indirectly, industries would be left with no informed basis to anchor their self-regulation efforts in. This will not only result in sub-optimal and ineffective efforts but also result in costs including wasted efforts and resources among others.
    \item \textit{Technology self-regulation- more AI to \textit{regulate} AI:} This argument emphasizes that we need more AI efforts to successfully come up with capabilities to guard against the risks emanating from AI. This is unfortunately a highly misinformed position to take. While there may be merit to the approach in a narrower context, it cannot be the entire policy position. 
    \item \textit{Regulating AI may put nation(s)/societies at disadvantage:} Another argument is that regulation will slow down advancements in AI and hence will yield the edge to national adversaries. This is an extension to the above argument suggesting that regulating AI risks killing innovation~\cite{sharma-2023}. A related version of this argument is that, even if there is broad consensus on the regulation and policies around AI, bad actors will have access to it and the only way to deter them is to have unregulated (or minimally regulated) AI. Again, this is not true as evidenced by technologies of the past. It is important and certainly possible to ensure robust innovation support while also ensuring safe, responsible and meaningful social implications.
    \item \textit{To regulate AI, the focus should solely be on the core AI technology:} There are two aspects of this broad argument. 
    \begin{enumerate}
        \item The first one is around guarding AI constituents - those of developed AI models and those that undergird this development. For instance, in the case of former, this includes model weights~\cite{WH2023}, and training data, while the latter include the hardware infrastructure such as GPU's~\cite{Schumer2020,mccaul2020}. It should be realized that AI is fundamentally different from technological advancements such as nuclear capabilities that it is often compared to in many ways:
    \begin{itemize}
        \item[-] easier access thanks to open sourcing, publishing practices (and these certainly makes scientific community stronger but come with challenges in the case of AI~\cite{seger-2023,harris-2024,owen-jackson-2024,doerrfeld-2024})
        \item[-] no fundamental barrier to entry given the development methods and underlying algorithms are relatively well-known
        \item[-] regulations against access to infrastructure can often be evaded (and in many cases are eased due to exceptions)~\cite{baptista-2024}
        \item[-] various indirect ways exist to access the technology (e.g., via investments, purchases of exclusive licensing, and indirect access to products such as ChatGPT)
    \end{itemize}
    Hence, while some of the efforts to guard supremacy can provide incremental and temporary advantage at best, they do little to help ensure that AI yields meaningful benefits and the threat envelop is reduced.
    \item The second aspect to this argument relies on narrowing the AI policy approach to the core AI development (and not application). It states that the AI policy should be predominantly centered on responsible AI development process -- for instance, by building guard-rails in the AI model themselves, ensuring fairness and absence of bias in the models, ensuring a kill-switch for the models if they behave in unintended manner, and to have interpretable and explainable models. While a subset of these aspects are indeed important and should be pursued, the argument unfortunately demonstrates a significant lack of understanding on both the AI technological front and also how it gets integrated, adopted and internalized in various areas. Without the utilization context, it would be impossible to build any robust safety measure in a balanced manner -- that is measures that ensure that the technology operates reliably \textit{and} effectively.
    \end{enumerate}
    \item \textit{AI progress would naturally result in positive societal outcomes:} This argument that technological progress by itself will lead to the right societal outcomes is not based on solid grounds. Further, the fact that scientific progress can be achieved in vastly different value-driven societies further emphasizes the fact that the societal value system should be based on much more than technology alone - it should come from a collective consensus on how we want our societies to be and for them to evolve into. These points can be seen evidenced in~\cite{wagner2024}. There is another context in which this argument appears specially troubling -- that of, assigning (implicit) 'altruism' to AI. Despite all the anthropomorphizing of AI, it should be understood that AI (at least as it stands presently) doesn't not have any underlying reasoning capability, let alone a core value system or agency. Consequently, the argument that AI will somehow match the value system of human society and optimize itself is misinformed at best and misleading at worst.
\end{enumerate}

Most of the above arguments do not have evidence or sufficient rationale to back them up. However, collectively they have served as a distraction crippling effective policy efforts.

\subsection{A fragmented piecemeal approach}

As a result of the various above factors, policy efforts have so far managed to focus only on primitive proposals and that too in response to specific challenges that have appeared from AI. While we should be prepared to address specific unanticipated challenges, such an approach for an area with a wide expanse and implications as AI requires us to work in a comprehensive, \textit{informed}, and forward-looking manner.

Unfortunately, most of the proposals, and a very small subset of it that has come to fruition as policies, is a consequence of looking into the rear-view mirror. That is, they have been \textit{reactive} in nature~\cite{ai-bills-2024}. One of the most recent example of this reactive policy-making is around deepfakes. \cite{graham2024} summarizes various steps taken both at federal level and at states' level in the US to deal with deepfakes issue in various context. It points out correctly that there is no comprehensive policy on dealing with deepfakes. Various piecemeal measures have arisen as a result of having to deal with specific situations. For instance, the Washington state bill focuses on sexually explicit deepfakes, while the ones from Louisiana and South Dakota focus specifically on minors, and New Mexico focuses on deepfakes' use for advertising.

Similarly, various responsible-AI proposals have appeared \textit{in response to} issues arising from the use of AI, mostly focused on GenAI. For instance, legislation such as~\cite{Berman2024} in response to disinformation and misinformation in the context of elections, and proposals such as~\cite{Clarke2023} in response to issues related to GenAI-created deepfake or misleading content and associated liability. Refer to~\cite{ai-bills-2024} for a list of various bills before the US Congress aiming to address concerns around AI. As is evident, most of the energy of legislative effort is consumed by GenAI related concerns and consequently we risk missing the bigger context around AI that impact society in various ways. \textit{And this risk of missing the real AI context cuts both ways - while missing the breadth of the AI risk landscape, we simultaneously risk putting in a potentially unduly burdensome, but ineffective, regulatory structure.}

The fundamental mistake is that of casting the issues above either purely as AI model problems (again, influenced by some of the factors influencing current AI policy regime discussed above) or as application problems in a very narrow context without accounting for associated mechanisms and dependencies.  In some cases, the argument that certain issues are properties of AI technology development process holds true. As we mentioned earlier, in the case of hallucinating GenAI models, the questions are not just how do we increase the accuracy of the models but also how do we ascertain that the utilization of these models in various contexts will deliver fair, responsible, and reliable offerings (products or services). Similarly, when guarding against deepfakes, efforts such as~\cite{Clarke2023} focus on intentional deepfakes and the associated liability. It should be noted that such actions can be very difficult to enforce barring in some clear intentional cases since much of the deepfake generation can be (and usually is) automated. For instance, it'd be extremely difficult to track whether there was an intentional deepfake generation or if it was an artifact of the underlying model used in an automated workflow.

On the AI practice side, most efforts have focused on adding guardrails along AI model development (e.g., transparency metrics, data usage, bias-awareness and interpretability). However, this doesn't guarantee that the models will be used in a responsible manner downstream. The undesired outcomes as a result of AI integration are not consequence of \textit{just} the AI model properties. Proper utilization of these powerful tools also needs to be ensured for fair and just outcomes~\cite{ross2023,viceuber2023}. Even when focusing on the effectiveness and risk of AI models, our current assessment approach to AI almost entirely overlooks \textit{reliability}.

Our current AI policy approach lacks coherency and more importantly fails to guarantee the outcomes that it strives for. This lack of coherency has resulted in:

\begin{itemize}
    \item[-] reactive piecemeal policy proposals
    \item[-] a relatively myopic casting of policy challenges as mostly technological challenges
    \item[-] multiple disconnected and narrow efforts leading to unmanageable and unenforceable policy apparatus
    \item[-] focus on methods instead of outcomes
    \item[-] treating AI in isolation and not in conjunction with other technological and social developments
    \item[-] confounding the risk posed by AI presently with those claimed at the ends of risk spectrum
    \item[-] no mechanism to correct for damages already done, e.g., not enforcing policy frameworks retroactively (thus benefiting reckless actors for their past actions)
    \item[-] ineffective democratization efforts on the opportunities across enterprise sector and consumers alike
    \item[-] lack of representation of the most impacted stakeholders in policy-making
    \item[-] lack of focus on disparate but accumulating/agglomerative risks from AI adoption (e.g., social engineering efforts through deepfakes)
\end{itemize}

The inherent limitations of this fragmented technology-centric approach are posing roadblocks in real-world implementation too. Let's look at a few examples:
\begin{enumerate}
    \item The recently announced NIST AI Public Working Group building on AI Risk Management Framework~\cite{nist-ai-rmf-2023,nist-ai-rmf-playbook2023} forms one aspect of this overall strategy. This is indeed a well-intentioned start with significant thought put into managing AI risk. However, it is relatively light on the actual methods, technological needs, and the current and future state of risk management. As an example, in its current form, it focuses on AI systems themselves but their use, esp. in combination with other AI and non-AI systems, can pose many unforeseen challenges. Further, there are no well established and agreed upon metrics or frameworks for measurements and mapping as recommended by this proposal. Finally, this framework doesn't address any adversarial scenarios –- scenarios where users can elicit unwanted behavior from these models, or intentionally exploit design security holes~\cite{nasr2023, brown2022, carlini2021, carlini2023, huang2023} -- something that has the potential to be a highly critical challenge.
    \item Innovation in detecting AI generated content to combat mis- and disinformation predominantly propose to use AI to combat challenges from rapidly evolving AI and are constrained by definition (e.g., detecting fake AI-generated content using AI techniques). Overall efforts for \textit{information provenance} (not just data provenance) need to be thought of and devised in fundamentally different ways.\footnote{ Information provenance goes beyond content provenance and authenticity establishment efforts (e.g.~\cite{c2pa2024}) and ensures that the entire \textit{information whole} is verifiable, and not just its constituent pieces such as included media and references.}
    \item Proposals such as the PACT Act~\cite{Schatz2021} include clauses to press companies that are building core AI technologies (e.g. GenAI foundation models) to make the data available for research projects with NSF approved grants. However, such provisions still leave out the risks that such technologies would pose given the lead time that the companies will have. For instance, the research community will certainly have a delayed start since such data will be made available typically long after these technologies have been developed by the companies and even potentially introduced as part of their products. In addition, it would be unrealistic to assume that the companies would also provide entire data in raw usable formats along with metadata and other supplementary resources used for technologies powering commercial products for a variety of reasons. Finally, it will miss all developments that would happen through private or corporate funding sources.
    \item Cyber-security efforts directed at safeguarding foundation models such as  the one required by the White House AI Executive Order~\cite{WH2023}) or that proposed in the RAND report~\cite{Nevo2024}, risk having limited benefits.  As more open-source models are being developed and made publicly available (e.g.,~\cite{llm360team2023,ai2team2023,meta2team2023}), guarding specific model weights would mean guarding a corporate asset not the underlying technology. Moreover, various foundation models will very likely converge in terms of their performance and hence having an edge just in terms of marginal performance guards against neither technology leak nor its misuse.
    \item The current policy thrust is mainly focused on safeguarding AI assets or building a risk management framework (mostly guidelines since not enough relevant metrics exist). It is much more important to address the ongoing impacts of AI on various current areas of societal and national security relevance.
    \item An effective policy needs to help incentivize the AI innovations of societal import. Hence, the technological development in core AI should be understood and addressed in the context of its applications and encouraged accordingly. This then needs to include all relevant areas including healthcare, strengthening democracy, societal safety architectures, privacy protection, protecting vulnerable users, impacts of AI-driven applications on social issues (e.g., addiction~\cite{dreyfuss2024}), societal impact on minors (e.g.,~\cite{dreyfuss2024}), children and youth related issues, education, civic awareness, and impacts on the efforts to build a well-informed populace. Consequently, such policy efforts will need to account for interaction and interdependence of AI with other technology areas such as social media or use by information-, media-, and education enterprises. Our current policy approach doesn't accomplish this goal. Some initial proposals on addressing a subset of the above issues are being put forward, albeit again only from a technological perspective~\cite{smith2024}. 
    \item The existing policy approach struggles in drawing connections between AI technological developments and their subsequent effects on the products, offerings and services that will impact and/or interface with various functions of the society -- be it social media, hardware products, software services, enterprise software powering various capabilities, national defense, cyber-security, media and information industry, education, healthcare industry, or automotive industry.
\end{enumerate}

The above list is meant to be representative and not comprehensive. With the current technology-centric approach, we will continue to see such instances expand inevitably leading to an unmanageable policy and regulatory setup that will pose more challenges than providing clarity.

\section{Constituents of a Comprehensive AI Policy framework}\label{sec:constituents-of-ai-policy}

\begin{figure}
\centering
\includegraphics[width=475pt,height=225pt]{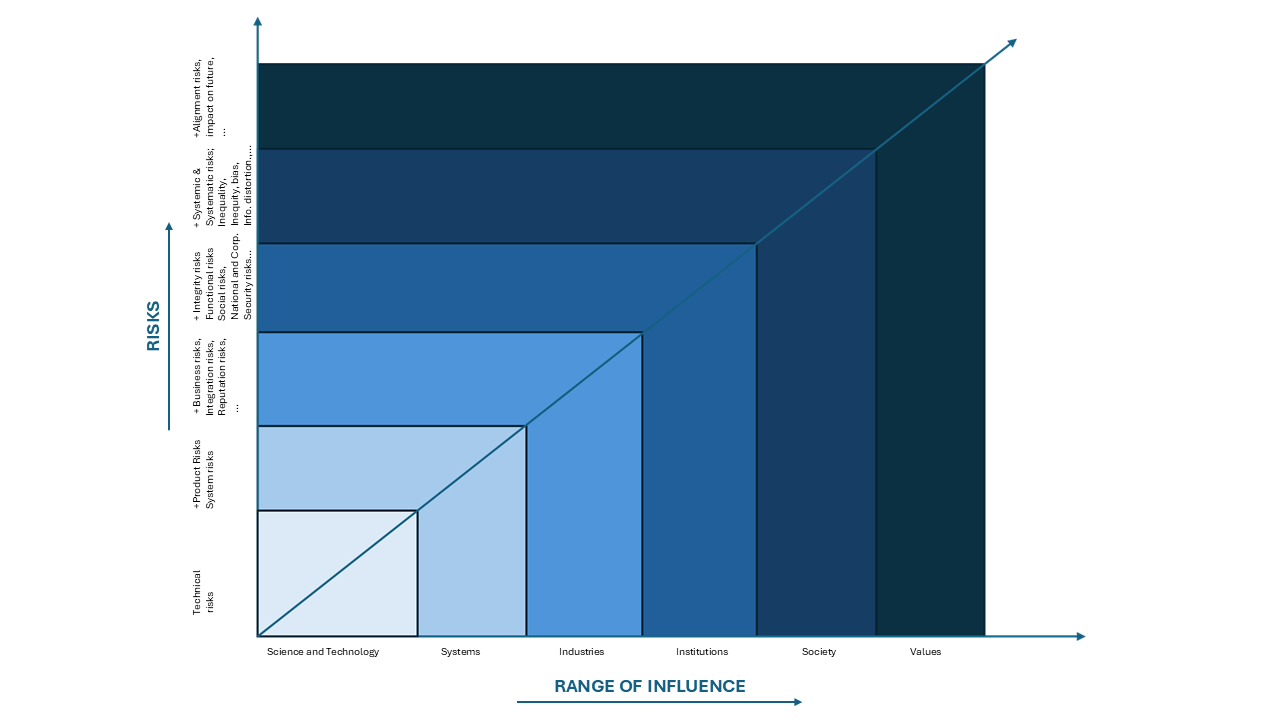}
\caption{AI Risks increase with increasing range of AI's influence}
\label{fig1}
\end{figure}

An effective AI policy framework would comprehensively and coherently need to address AI development, integration and adoption across various realms and reconcile them with the existing regulatory and policy frameworks.  It is hence important to understand the basis underlying the reach of the AI technology and its interaction with various aspects of society. As Fig.~\ref{fig1} shows, the risk vectors grow significantly as the technology's range of influence expands across increasingly widespread functioning of society, both directly (as consequence of AI-driven products and services) and indirectly (as the downstream effects of adoption of AI-driven capabilities). We are already witnessing an increasing penetration of AI- and data-driven capabilities in various facets of our daily lives from weather prediction, navigation, search to entertainment and information sharing. Further, the concurrent technological developments allow for rapid introduction of new AI-driven products in various areas and domains -- day-to-day (e.g. social media) and specialized (e.g., understanding of biochemistry, drug discovery, legal, and even scientific research itself). Clearly, as the technology demonstrates more promise, its adoption is growing rapidly and along with it the risk spectrum is expanding significantly. Moreover, the time between the introduction of new technological developments and their adoption continues to shrink due to various factors discussed in Section~\ref{sec-intro}. 

Fig.~\ref{fig1} provides then an important understanding of the nature and gravity of the risks as the technology penetration progresses in various areas. The horizontal axis in Fig.~\ref{fig1} shows the increasing AI penetration across various areas in the order of importance. It should not be interpreted as a sequential graduation of technological influence. Developments in various applications of technology allow for different areas to be impacted in parallel, or even out of order. For instance, a new social media application can have an immediate impact on the society proportional to its adoption. Such applications are rapidly introduced directly to the users and almost never follow any rigorous testing and validation. One of the reasons that the guardrails aren't adopted is that there are none required in the absence of a social consensus, policy guidelines or regulations. Only much later are the challenges realized. In the case of social media powered by AI, we have mentioned various challenges arising from information distortion, scams, exploitation of women and minors, to the impact on mental health among others~\cite{striped2024,kristof2024,chen2024,dreyfuss2024}. Similarly, GenAI offerings, powered by LLM's, have shown an array of challenges across different areas~\cite{lecher2024,wetsman2022,toulas2024,smith2024,swan2023,haim2024,graham2024,katz2024,koebler2024,goodin2024,hicks2024,robey-2024}. These challenges keep evolving and the new ones continue to be discovered as an almost uncontrolled adoption of a promising but non-mature and/or untested technology continues at an accelerating pace.

It is evident then that the current reactive AI policy approach is a losing proposition. The policy framework needs to stay ahead of the technological developments and have a more informed platform to guide these developments in socially beneficial directions while minimizing the risks, and also have mitigation strategies for the risks that do realize.

Armed with these insights and the risk spectrum of Fig.~\ref{fig1}, we can identify some key priorities that should constitute the foundation of AI policy framework along the dimensions of society and value alignment, institutional readiness and preservation, national and corporate security, technological safety and reliability, and future readiness:

\subsection{Maintaining and Reinforcing the Social foundations, Institutions, and Democratic values}

Just like any policy efforts, the fundamental construct and basis of an AI policy should be to ensure that the core social foundations and democratic values are maintained, and ideally reinforced. An effective policy would focus on guiding the technology (or any other policy objective area) to reinforce and strengthen these pillars. At the very least, an effective policy framework should make sure that the introduction of technology in various realms of social life does not erode these building blocks of society in any manner. Hence, the policy's top mandate should be to ensure the maintenance and reinforcement of the core social constructs including democratic values, institutional integrity, equitable benefits, just outcomes, respecting and maintaining constitutional rights, fairness and transparency, equal opportunities, maintaining human dignity and independence, participation, and objectively informed citizenry. All the specific elements of an AI policy framework should follow from this core objective.

\subsection{Ethics and \textit{Alignment}}

With expanding usage, the outputs from AI-powered systems and products will be further implicated in various areas. Respective ethical and risk frameworks that account for such incorporation and adoption will also need to be built and assessed. Examples include drug-candidate proposals from AI systems, AI-derived insights for policy guidance on areas affecting swaths of population, and AI-derived target identification or potential population surveillance in war scenarios(~\cite{humble-2024,jinsa-report-2021,frenkel2024,pfau2024}). Such outputs and systems need to be seriously vetted, and the risks addressed before their use directly or indirectly. Addressing such challenges will go beyond just evaluation or validation of AI systems to ethical and responsible use of resulting technologies and hence placing guardrails on the \textit{use} of AI systems. Many other areas are being impacted by AI use and it is important to discuss not just ethical use but also to align on whether the use is desirable in the long-term interests of the society (see, e.g.,~\cite{coleman2024,haim2024}). This latter point is indeed a core thrust of the policy framework proposed in this paper. While there have been some isolated attention given to the \textit{alignment} question for AI in its current and potentially future forms, these discussions are incomplete and even distracting in some cases. There have been discussions of AGI (or ASI) to denote a future stage of AI where it can achieve human-level capabilities in reasoning, planning and agency and such superintelligence potentially entirely replacing human decision making. This hypothetical scenario currently forms the basis of vague efforts to achieve some type of alignment for these potential models with human values. Also, the efforts claim to apply the alignment efforts' outcomes to current incarnations of AI models that are decidedly nowhere close to human-level reasoners and planners. While alignment can be considered a worthy goal, it is by design unattainable as it is approached currently. There are various reasons for this. While, we do not delve into the details and arguments on this dimension, suffice it to say that the two main reasons why this approach is likely to not succeed are:
\begin{enumerate}
    \item the current predominant purely scaling-centered approach to AI (via. LLMs) is almost guaranteed to not be a pathway to AGI. More importantly, it'd be premature to predict what type of "intelligence" can and will be achieved in automated systems and what would its governing principles be; and
    \item the fundamental coherent basis to inform the alignment effort does not exist (or at least agreed upon), in the absence of which, no informed alignment work can be performed.
\end{enumerate}

Most importantly though, the issue with this alignment approach, ongoing discourse and the investment in it is the associated opportunity cost. In the foreseeable future, we will most certainly be dealing with AI models that will drive products as enabling technological components the \textit{use} of which will decide their effects and implications on the society. Hence, to achieve equivalent of alignment, we need a robust anchor in desired social outcomes, priorities and consensus value-system that both guides and guards these AI-driven offerings \textit{presently}.

\subsection{Systemic effects and just outcomes}

There can be both intended and unintended adverse consequences of AI adoption in driving business outcomes. This can have a significant undesirable impact on society including consumer choices, labor and workforce disruptions, and wealth and opportunity inequality (see, for instance,~\cite{acemoglu2024}). While some of the systemic risks resulting from social media in the form of resultant mis- and dis-information, effects on the education, growth and mental health of the youth have been highlighted above, there can be other underappreciated aspects that can pose social and societal challenges such as algorithmic wage discrimination~\cite{dubal2023,wray2024} and user-discrimination~\cite{ross2023} and -exploitation~\cite{viceuber2023}. With increasing use and adoption of AI across the board, the risk of such undesired effects and outcomes has the potential to grow significantly as well, and beyond just business applications. It is very important that the AI policy has embedded mechanisms to be able to track, address and mitigate such outcomes. A systematic approach to understand and alert for such systemic effects can lead to proactive policy framework to guide AI adoption.

\subsection{AI Robustness and Reliability}

Existing efforts to assess the robustness of AI products and services are relatively primitive, with some exceptions for non GenAI cases in the context of safety- and mission-critical systems. While there is significant discussion, seeming progress and an artillery of tools for evaluating, monitoring and safeguarding AI models, there is little of relevance when these are not anchored in the outcomes resulting from the application of AI models for various uses. This is most easily visible in the context of GenAI models almost irrespective of the domain that they are applied to. While a lot of discussion and investment is devoted to ideas that aim to address the inherent issues with LLM's such as hallucination and lack of reasoning (e.g., Retrieval-augmented-generation or RAG~\cite{lewis2021}, applications of ideas such as chain-of-thoughts~\cite{wei2024}, incorporation of Reinforcement Learning techniques~\cite{wen2024} and so on), little progress has been made in any robust terms to establish the reliability of either the models and/or the supported products. In fact, reliability in the current application of GenAI is almost an after-thought. Even the safety of these models can't be established let alone guarantees on their performances. Moreover, given the obsession with scaling-centered approach to build more sophisticated GenAI models, little if any progress is being made on building a theoretical understanding or rationale for the area. Naturally, we continue to see novel issues coming up with both the new GenAI models~\cite{xu2024,goodin2024,walsh2023,haim2024,tiku2024,lecher2024} and also (the non GenAI) traditional AI driven products~\cite{robertson2024,striped2024,coleman2024,pfau2024}. Not all these issues can be handled simply by establishing the behavioral predictability of the AI models. Much more effort need to be placed to perform testing and validation beyond stand-alone AI technologies extending to systems, system-of-systems, products and services in \textit{specific context of applications} and \textit{along metrics of actual relevance}.

Most of the proposed AI readiness assessment, esp. for GenAI, are either lacking in context (e.g., testing against benchmarks) or playing a game of catch-up as a post-hoc effort to reign in the undesired outcomes (e.g., detecting deepfakes). We need to move beyond this to more rigorous \textit{context-sensitive} testing, validation and verification regime for both core AI technology and AI-driven products with special emphasis on \textit{reliability}.

This needs a change from bottom-up technology-first approach to an outside-in society-first approach. For instance:
    \begin{enumerate}
        \item Efforts in assessing increasingly complex and sophisticated AI capabilities need to be anchored in business and/or social functions they serve as part of products and services, moving beyond current approach of contained and limited evaluation. The utilization of technology and associated product context should determine and guide the risk management strategy. Frameworks such AI-RMF~\cite{nist-ai-rmf-2023,nist-ai-rmf-playbook2023} will be necessary but not sufficient in such scenarios.
        \item For critical technologies and applications, we need to move from run-of-the-mill evaluation to robust \textit{validation and verification} mechanisms.
        \item The testing and validation efforts should expand from AI algorithms to overall application systems and their utilization. For instance, validated AI algorithms can still have potential to be employed in a manner that has adverse implications on public interest~\cite{ross2023}.
        \item Tractable risk management mechanisms for enterprise use and implications, public-facing exposure of products and services, safety critical systems and mission critical systems need to be established. Existing frameworks can also be expanded to account for AI subsystems that will be added or replaced, owing to their stochastic nature. For instance, automotive are regulated in terms of their public safety standards esp. since they pose risks to life. Similarly, hardware robots have safety protocols that they adhere to. Different domains of applications will need respective safeguards aligned with their criticality. These can be addressed with respective agencies in partnership with AI policy bodies. For instance, NHTSA can have an expanded mandate to cover safety of AI systems as part of the transportation safety. Further, potential use and adoption of AI in areas such as defense and weapons program (CBRN -- Chemical, biological, radiological, and nuclear) need to be further validated both for safety and for unintended consequences.
        \item Risk management frameworks need to be expanded to cover system-of-systems as well as the up- and down-stream implications. For instance, the social media content that is served to the users via recommendation algorithms (e.g., on instagram, or tiktok) can have an impact on users' behaviors and also have broader societal implications when used in conjunction with algorithms that prioritize content of highly sensational nature, false information or propaganda. Personalization algorithms for content serving (e.g., news, products, media content) can result in challenges -- the upstream risk comes from having continuous personalized profiles of the users making them susceptible to tracking and loss of privacy, while downstream risks can range from risks of social engineering, anti-competitive behaviors, and reinforcing biases. 
        \item With AI either powering or becoming subsystems of various hardware technologies (e.g., automotive, industrial systems), it is important to also manage the safety implications of such hardware and industrial assets.
    \end{enumerate}

    \subsection{Accountability and Mitigation Framework}

    As the AI adoption has picked up pace, the learned AI models, the underlying data used to train these systems, the products and services powered by AI as well as the downstream implications are posing new questions and concerns on the legitimate use, privacy, IP infringement, societal costs, risks and potential for harm. A variety of concerns have already been put forward from areas as wide as autonomous driving, social media, defense, finance, social inequality, and wage discrimination among others. As these risks become more apparent, we still lack a clear mechanism to address these concerns, establish and enforce accountability and take mitigating actions to correct the wrongs. This in part can be attributed to the rapid pace of AI development and integration that has taken the policy and regulatory structure by surprise, the lag in the legal framework to catch up, and a lack of comprehensive understanding of the AI landscape. In the areas that these concerns are concentrated (e.g., news, media and creative workforce, and authorship), the respective communities are trying to have them addressed through various mechanisms including legal action~\cite{ai-lawsuits-2024} and explicit agreements with AI companies on data use agreements~\cite{david-2024}. Lawmakers are also trying to propose various bills in reaction to the apparent damages in some cases~\cite{ai-bills-2024}. However, most of these actions are either piecemeal or very specialized -- thereby focusing on narrow aspects of the issues while ignoring the root causes and systemic effects.

    We need a policy framework that has clear mechanisms to address the accountability for the risks introduced by AI-driven capabilities that are meaningfully associated with the desired outcomes, not decoupled from them. An accountability and mitigation framework should also address others needs such as:
    \begin{enumerate}
        \item Have a robust liability framework delineated across parties from entities building foundational models, entities integrating and refining them for purpose-built systems, and products and services integrating them. For instance, if a product powered by a commercial foundation model ends up compromising user-data from prompt engineering (a security gap demonstrated by research for foundational models~\cite{fu-2024}), the liability may need to be extended all the way to the foundational model provider. In other cases when the misuse of AI products has been intentional by other actors (e.g.,~\cite{viceuber2023,ross2023,wray2024}), the framework should be able enforce liability on such business users.
        \item Include rigorous policy on sharing, transfer, and export of core and supplementary technology.
        \item Build a collective framework to establish the safety and security of technology (e.g., security holes and data compromise probabilities). Such a framework will need participation from both industry and regulatory authorities. Some initial efforts have been made through such a collaboration resulting in the ISO/IEC 42001 standard for AI Management System that can provide guidance. However, such standards, adoption and uniform practice framework need to be further reinforced and strengthened.
        \item Having clear protocols to ensure adherence to intellectual property rights and avoid/penalize violations (e.g., permission to use public data, protected data, copyrighted data).
        \item Validating the correctness of outputs/outcomes to minimize mis- and dis-information in both original and derivative forms (e.g., outputs being used by other subsequent systems as inputs).
        \item Include clear policy on advertised claims esp. for technologies of public import when introduced in the product -- something akin to a safety warning label and/or nutrition information for food products -- informing users of limitations and potential risks of using the technology, available in an easily accessible and understandable format. 
        \item Having framework for \textit{graduated testing and release} for sensitive application areas. Graduated testing based on agreed-upon 'safety-markers' (akin to drug approvals) can allow for well managed public-release of safety critical products and services.
        \item Incorporating clear mechanism for product  \textit{recall} upon discovery of potential risks and have guidelines on addressing the issues, re-validating and verifying the products before reintroduction.
    \end{enumerate}
    
\subsection{Future readiness - R\&D and Innovation strategy}

The evolution of AI technology can't be reliably predicted, and we should be prepared for novel and expanded risk vectors. So far, most attention has been paid to immediate characteristics of the AI models with very high emphasis (and arguably rightly so) on the emerging GenAI landscape. However, we'd be remiss if we do not contextualize this with the concurrent and potential technological and social developments. We need to understand the risk-reward trade-off (see Fig.~\ref{fig1}) in this evolving landscape as we make collective decisions to incentivize, adopt and safeguard various developments. This necessitates an effective information mechanism that allows to surface not only the potential emerging risks and challenges, but also high-priority opportunities that can help address the highest order needs of the society. After all, the goal of any effective policy framework is not only to reduce unlawful and damaging consequences but also to promote and accelerate responsible developments for social benefit.

This is where a proactive policy arm that actively promotes meaningful research and innovation strategy becomes important. It needs to identify both the priority areas of innovation and also the mechanisms to enable them. A segment of the scientific community has been warning against the potential existential risk of AI and advocating for some immediate thrust in relevant research areas (see, for instance,~\cite{bengio2024}). However, it is clear that the warnings and proposals to address them do not necessarily match. Further, the warnings pose a hypothetical future (mostly scenarios imagining availability of some form of AGI without a clear consensus definition of what it would be) while the risks, as we presented in this paper, are much more amorphous, immediate and already manifesting themselves.  

It is critical that the national innovation infrastructure be examined for readiness to deal with both immediate and future challenges, be reinforced to focus on emerging challenges and opportunities along with technological and social evolution, and readied with an effective execution mechanism across various areas of national support and incentivization.

Fresh approaches to address the growing risk of AI should also be encouraged and enabled. The status quo of doing incremental work using existing methods will certainly be insufficient. For instance, detection of AI generated content is important but will always stay behind the newly developed techniques to generate synthetic content. Moreover, developments in such AI-generated content detection approaches can themselves serve as a catalyst for novel AI algorithms that can trick such detection systems. One such novel approach for establishing information provenance can be a \textit{trust-based internet design}. Some initial efforts in establishing media content provenance and authenticity are being developed collaboratively in the industry (Coalition on Content Provenance and Authenticity -- C2PA)~\cite{c2pa2024}. Efforts such as C2PA can be used to bootstrap broader initiative towards developing a holistic framework.

\subsection{Addressing Externalities}

The rapid AI development and adoption also result in a variety of externalities imposing direct and indirect costs to the public. A comprehensive policy framework should develop mitigation plans as well as incentivize innovation to account for, and counter, these externalities. Some examples of such externalities include:
\begin{enumerate}
    \item Use of user-generated data without consent: Once done, such actions are almost irreversible and hence the public has very limited recourse upon non-consented data-compromise.
    \item Sustainability impacts: Increasingly complex AI models such as those in GenAI are extremely resource intensive to build, maintain and adapt. Such developments are slated to further accelerate with an increasing potential to impact natural resource footprint (see, for instance,~\cite{c2pa2024,ren2023,belanger2024,galaz2021}). The rapid pace of developments in the AI landscape are already pushing for solutions to address the energy demands with relatively minimal oversight and public engagement on the choices made.
    \item Consequences for society: AI-driven products and functionalities can have serious implications on areas such as dialogues and discourses impacting the fabric of society, data privacy and security, vulnerable populations, financial stability and income inequality  (see, for instance,~\cite{kristof2024,goodin2024,danielsson2024,maham2023,acemoglu2024,striped2024}). It is important that the policy efforts have a continuous $360^o$ visibility to such current and potential consequences.
    \item Proposed methods to deal with AI risk can themselves pose challenges. For instance, the use of user-verification and identity association with internet activities to establish user authenticity, also pose the danger of privacy loss and compromise of safety and security (e.g., by highly granular user-behavior exposure, and tracking capabilities) and can also provide tools for surveillance. As an example, to avoid bots or automated agents from using ChatGPT API’s, OpenAI requires users to be verified. Similar efforts have been adopted in the context of avoiding fake contents from unverified sources or use from adversarial actors. However, such approaches may result not just in the loss of privacy of the users, but also opens up risks resulting from compromised information and/or modeling personal behaviors. A more recent example in this direction is the so-called \textit{proof of humanness} achieved via dedicated devices across all arenas of our daily lives~\cite{world-2023}. Such efforts should be rigorously examined, and potentially regulated if necessary, before being adopted.
\end{enumerate}

\subsection{Addressing Adjacencies: Related and concurrent technologies}

Multiple concurrent developments in technology in various areas jointly shape our world and our society. As such, the effects and implications of AI adoption also relies on the intersection and interaction of AI with other technological developments that do not just impact each other but taken together can, and do, impact various areas in profound manner. It is important that the AI policy framework accounts for these ongoing developments in critical technologies including semiconductor manufacturing, chip design, quantum computing, IoT, blockchain, software, social-media, communication and cyber-security to name a few.
%– have the potential to pose further challenges that should be understood and efforts towards addressing them initiated.
Consider how advancements in data technologies when taken together with AI can pose complex challenges. Advances in data technologies allow for gathering, transforming and processing of vast amounts of data at scale. Data from various realms, both independent and copyrighted, are becoming increasingly common and routinely inform AI technologies' developments. New business models are being developed by companies around selling user data to AI development firms (e.g.,~\cite{david-2024}). A swath of data already used to train GenAI models has raised serious concerns from various impacted stakeholders (see, for instance,~\cite{ai-lawsuits-2024} for a list of copyright and other lawsuits in the context of AI training data) and can have serious implications for entire industries and workforce. 

To make the above problem more complex, consider the effect of connected products that surround us in a variety of forms including 'smart'-products (e.g., thermostats, wearables, consumer appliances, voice-assistants, phones and even vehicles) that have the ability to capture real-time data and activities of users. This takes the above problem of user data to the next level -- it is not just the private and/or proprietary data but also behavioral data streams that get exposed for various applications and uses. While such data can be used to improve products and services, it can just as easily be used to build people profiles, tracking information, lifestyle and behavioral profiles and so on the use of which shall be well understood, monitored and collectively agreed upon. 

The rules and guardrails around data collection, usage, and further monetization by businesses is weak at best. In most cases, there is no precedent on companies choosing to monetize user-data shared for specific business purposes in non-agreed upon ways.  A strong data privacy and user-consent framework needs to be put in place to govern the use of such data by respective businesses and whether and how the service providers can build data-businesses by selling user information. This is important not just from a privacy perspective but also potentially from a national security perspective (e.g., when such data capture or data marketplaces enable foreign products and services to build profiles of US users). As an example, building such user profiles can pose misinformation and disinformation risks via social media, cyber-security risks, as well as risks to consumers and employees through potential abuse by businesses resulting in unfair and/or exploitative business practices (see, for instance,~\cite{viceuber2023,david-2024,dubal2023}).

AI can also accelerate the effects and risks associated with other technologies. An immediate example would be social media. Mis- and Dis-information are and should be one of the core themes to understand the direct and indirect impacts of AI. For instance, the problem is not just the fact that AI can generate deepfakes or mis-information content, but also in how AI can be used in \textit{propagating} this content. The mis- and dis-information content doesn't always have to be \textit{generated} fake content, but can just as easily be (and typically is) a spin on real content, partial and context-less sharing, and/or intentional mis-characterization of events. This phenomenon is rampant on social media and a big part of challenge is also in the underlying AI algorithms used to pick the kind of content that gets shared and promoted. It is such context with existing technologies such as social media that need to be well understood and accounted for. As mentioned earlier, such outcomes from AI-generated or driven content and priorities affect a range of industries~\cite{reyes-2023,trattner-2021,chin-2023,morrison-2023}, not to mention posing threats to institutional integrity and social harmony~\cite{robertson2024,acemoglu2024,Berman2024,Clarke2023,graham2024}.

In addition to other technologies, AI has the potential to impact entire realms of social life. For instance, incorporation of AI in technology offerings is poised to impact education and information realms which will have further impact on labor, workforce makeup as well as on the broad participatory nature of democracies. It is important to understand then the impact of AI powered products and services such as those in media, information-, ed-tech industry, and academia (see, for instance,~\cite{morrison-2023,reyes-2023,chin-2023,trattner-2021,ghafarollahi-2024}). More generally, impact-themes such as application-automation in conjunction with AI can affect large swaths of industries, application value-chain, and workforce. 

\subsection{National and Corporate security}

As AI technology gets implicated progressively in applications with national security implications in defense and warfare, critical infrastructure, public administration workflows, utilities, energy, finance, and more, it is critical that we have visibility and safeguards against both the technology failure and adversarial risks. Similarly, AI can not only mean competitive advantage in the corporate sense but also is becoming integral in corporate offerings that powers and run capabilities with national security relevance such as healthcare, benefits provisioning, navigation, search, digital news and information, and media. Hence, as the technology adoption increases, the policy framework needs to be well informed on the inherent risks of such adoption and adversarial challenges to ensure national and corporate security. 

For instance, as the AI models have become increasingly complex -- they can generate synthetic content, and can be queried directly -- the risk profile resulting from integration of these models to various systems have changed significantly and continues to evolve in relatively unpredictable ways. While there is potential for these models to be used in nefarious ways -- e.g., generating and disseminating mis- and dis-information~\cite{hicks2024}, performing automated transactions, build automated workflows, and emulate users -- there are also security risks resulting from the ability to query and direct these models to reveal confidential information~\cite{goodin2024}. These security holes can offer opportunities for malicious actors and can compromise sensitive information ranging from proprietary commercial and IP information to sensitive information on users including US citizens and from our allies. Some such challenges are already presenting themselves~\cite{chen2024}. This risk extends through all the assets be they public or private assets. Commercial enterprises lack readiness for managing and dealing with such vulnerabilities. In fact, recent research suggests a more profound risk landscape for generative AI ecosystem showing that adversarial actors can develop malware to exploit the generative AI components of agents and use that to launch cyber-attacks on the entire GenAI ecosystem~\cite{cohen2024} thereby using them as gateways to broader national infrastructure. Similarly, evidence is already emerging around the security risks that can appear from external or even open-source AI models used as a backdoor for malicious software posing both security and espionage risks~\cite{toulas2024,owen-jackson-2024,seger-2023}. This risk also applies to how the technology is developed, deployed, provisioned to the users, the safeguards to ensure that the IP isn't plagiarized or copied~\cite{Bastian2023,Heath2023}, and is not exported to adversarial actors. 

While risks emanating from the introduction of GenAI are already evident and pose immediate challenges as briefly discussed above, there are many other latent risk vectors that may have gone or are being overlooked. These include aspects like systemic issues arising from concentration risks in suppliers and providers, reducing independence of defense and critical infrastructure providers and introduction of indirect non-obvious dependencies. 

Since these threats are posed not just to the US but also to our allies, it is also extremely important to build strategic partnerships with close democratic allies to address them. Consequently, \textit{building international consensus and value-aligned planning and execution should be an utmost priority of the policy framework}.

\subsection{Up-to-date Legal framework}

There are myriad other cases and ways where AI and data technologies have been used resulting in troubling outcomes (see, for instance,~\cite{lecher2024}). Further, given the scale and ease of use of these technologies, such intended and/or uninformed misuse will exponentially grow. There have already been troubling cases such as those arising from easily available deepfake technologies~\cite{kristof2024} or selective or potentially improper use of big data tools~\cite{wetsman2022} (note that application of LLM’s with search constraints such as RAG’s can easily deliver technically “justifiable” but misleading outcomes, for instance~\cite{magesh-2024}). While some efforts on specific aspects are being taken up by governments (see, e.g.,~\cite{swan2023}), the problems arising from potential misuse of AI are non-trivial and expansive. It is important that the legal and regulatory framework catches up to this set of issues. Just placing guardrails around technology will not suffice. Self-evaluating AI would almost certainly not be enough if at all feasible. External checks and balances informed by social priorities need to be put in place just as is done in other areas. Accountability should also be an important part of this discussion and should extend to the spectrum of industry and consumer players much like any other safety-critical products and services as discussed above. One of the important elements of such a legal-framework would be a consensus on the nature of outcomes. In many, if not most, cases, AI is an enabling technology. That is, the \textit{functionality} that AI support will need to still be held responsible for the intended outcomes. Both the legal and regulatory frameworks in various domains need to be updated and adapted to account for potential impact from the AI functionalities. For instance, Section 230 of the Communication Decency Act ~\cite{csrreport2024} has governed most of the internet and the responsibility for the content on the internet has typically been that of the content generator. However, with GenAI driven functionalities, it is unclear as to how the content generating algorithms should be treated. Similarly, the liability of risks posed by any type of AI algorithmic decision making that replaces humans is not clear in most areas. It is important that the legal and regulatory framework is updated to account for the current state of AI development and is also ready for future developments. Doing so from a pure technological perspective will always prove to be lacking. As more issues come to the fore around misuse of AI (see, for instance,~\cite{tiku2024,katz2024}), the legal issues around AI will surface and pose novel challenges and risks~\cite{long2023,walsh2023,flynn-2024}. 

\subsection{Unbiased basis and platform to inform policy}

There are many sources of data, empirical studies, private and public sector stakeholders, vested interests, and even global events, macroeconomic and geopolitical realities that inform policy making. It is extremely important then that the policy making has access to non-partisan objective verifiable sources of evidence. This information platform should be unbiased, transparent and open to public scrutiny which will not only inform the policy making in a principled manner but will also be instrumental in building the trust of citizenry. The effort should certainly have participation from various sectors and stakeholders both direct and indirect. It must go beyond the current practice of mostly including AI practitioners and researchers along with the respective chosen AI industries and extend to experts who can complement putting together a holistic picture of AI's effect on the society. These will include social scientists, ethicists, climate scientists, respective domain experts, public representatives and more. The platform should objectively and dispassionately present the evidence-supported state of technological development, relevant opportunities, future potential, risk landscape, assessment of the effects of policies and evolving landscape of the intersection of technology and society in an actionable manner.

\section{A Social-outcomes and priorities centered (SOP) framework for AI Policy}\label{sec:sop-framework}

\begin{figure}
\centering
\includegraphics[width=350pt,height=175pt]{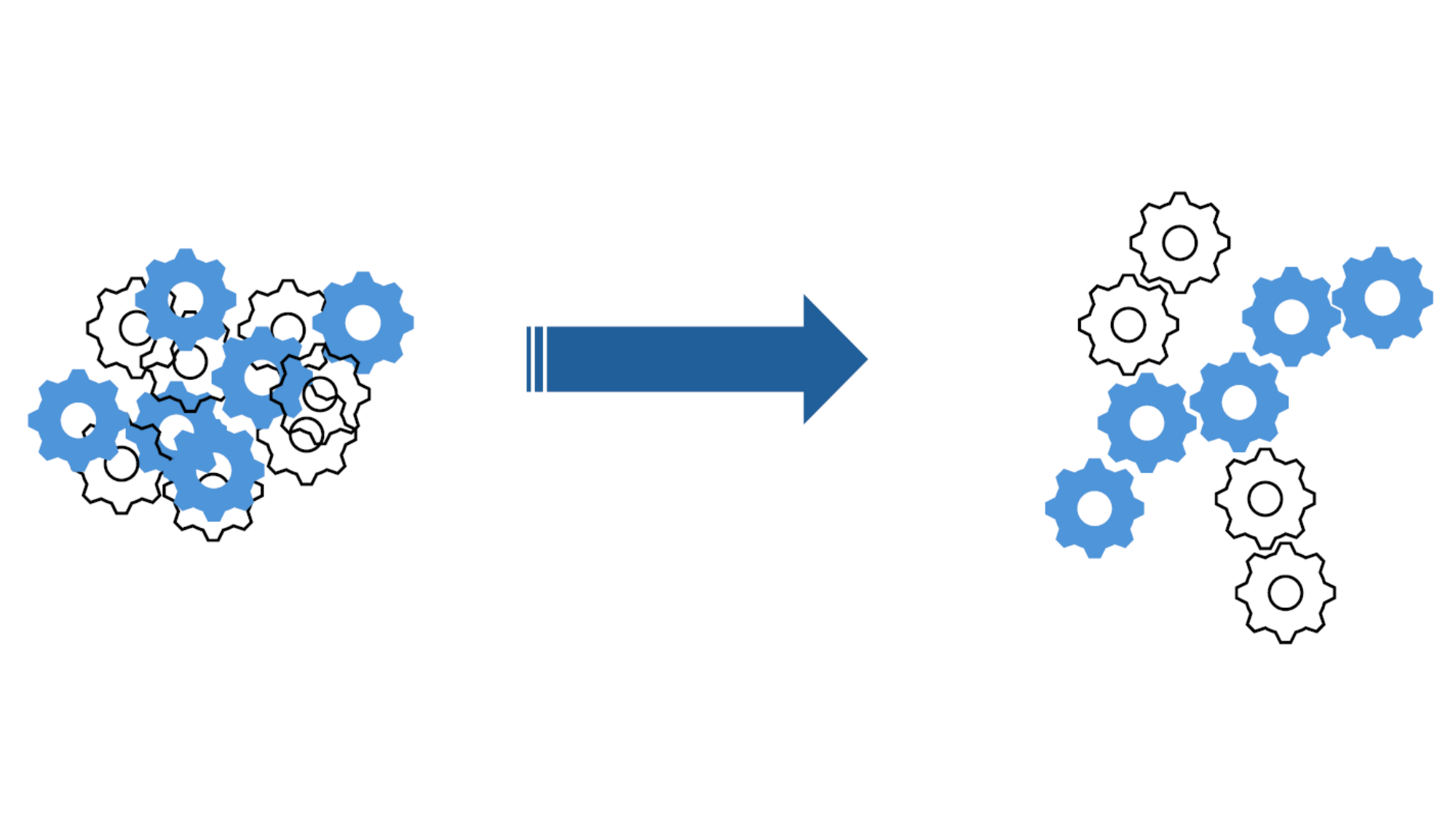}
\caption{Transforming approach to AI policy: From fragmented efforts to a systematic framework}
\label{fig-sop-framework-motivator}
\end{figure}

Naturally, it can seem daunting to develop an effective policy framework that encompasses the aspects around AI detailed in Sec~\ref{sec:constituents-of-ai-policy}. Building an effective and comprehensive framework requires us to pivot from the current fragmented and reactive policy approach focusing mostly on minimizing unintended consequences to a systematic \textit{policy framework} that \textit{also} aims to achieve desired consensus-driven social outcomes (Fig.~\ref{fig-sop-framework-motivator}). That is, \textit{instead of a technology-centered approach, we need a society-centered approach}. Interestingly, other AI policy efforts such as the ones by OECD are already witnessing such need for policy synergies along with an international call for cooperation on the technical aspects of AI, data, and privacy~\cite{oecdaiexpertgroup2024}. To this end, we propose a Social-Outcomes and Priorities oriented (SOP) framework for AI Policy. 

\vspace{1mm}
\textit{An SOP framework provides a mechanism to build policy and regulatory structure encouraging outcomes that strengthen society's core foundations and offer uniform and equitable benefits while minimizing the risks. It is a proactive forward-looking framework that doesn't focus solely on preventing harmful outcomes or mitigating them, but also actively promotes desired consensus-driven outcomes through appropriate incentivizations.}
\vspace{1mm}

The SOP framework aims to function through collective efforts of stakeholders from society. It is an inclusive framework that brings representation from \textit{all} strata of society -- first and foremost whose lives will be most impacted but lack enough representation and say in the discourse, government and regulatory bodies, academia and industry. A diverse set of input-providing stakeholders are critical to ensure that the proposals are well-informed both in terms of social priorities and also technical- and application-domain depth.

Informed and guided by actionable social priorities, such a comprehensive policy framework would enable responsible development, deployment, adoption, and integration of AI capabilities in products and services to optimize societal benefits while balancing innovation, regulation, accessibility, and national security interests. Moreover, when informed by social priorities along with the operational guidance, the policy framework would also stand the test of time and will be able to better accommodate the ongoing and rapid developments and changes in the AI technology itself. This is extremely important since the field is undergoing a constant evolution thanks to very high level of interest, effort and investment across the board. The EU AI ACT~\cite{eu-ai-act2024} is an immediate example of such a challenge to policy making. When the efforts on the EU AI ACT were initiated, developments in the GenAI area were not anticipated or foreseen. Hence these developments were not accounted for in the earlier iterations of the proposal. 

Finally, by pivoting our focus back to desired outcomes and priorities, an SOP framework would also allow us to target blind-spots when proposing policies. When starting with a technology-centered approach, the implicit assumption is to assign causality for any set of target areas to the technology in question. For instance, the mis- and dis-information on social media has often been attributed to issues with AI, or more broadly technology, alone (e.g., via deepfakes, or generated content). However, understanding the problems outside-in allows us to gain a fuller context. For instance,~\cite{robertson2024} demonstrates how most of the toxicity, conflicts, and fake news on social media is attributed to a very small minority of users ($3\%$ of users, $1\%$ of communities, and $0.1\%$ of users, respectively) but the disproportional impact that these sources have on the social discourse. An SOP framework would allow to focus on meaningful priorities -- for instance information reliability in this case -- and allow to address these issues holistically (addressing intentional non-technology assisted misinformation, industry standards for code of conduct in communities, regulating algorithms that reward content based on incorrect information, holding users and groups responsible, \textit{in addition to} AI generated content and algorithms aiding generation and dissemination of mis- and dis-information). This will allow the technology aspects of any legislative approach to be contextualized within the broader set of policy efforts to tackle the relevant issues and achieve desired outcomes. 

Often the discussions and proposals around AI policy are positioned as issues of \textbf{trust} in AI. The need for trust in AI has often been mentioned, emphasized and characterized as a north star of sorts. However, we have \textit{no consensus definition of this trust} or even a robust understanding of what the trust would look like and how it would manifest itself in practice. 

On the technology end of the trust spectrum, the topic is often inundated in technical approaches to address specific technical aspects of AI model building, evaluation or deployment. Examples of such approaches include a plethora of proposals such as responsible AI (again, a terms which itself lacks a definition), interpretable or explainable AI focusing typically on "understanding" how the models reach their decisions, evaluation to ensure models behave in certain ways (often done via benchmarking against a specific set of objectives, only empirically established at best), expanding approaches to constrain models from providing incorrect information or outputs (hallucinations being the prime example), and building models to guard against issues such as bias and deepfakes (either via integrating to model development itself or as post-hoc approaches). 

On the other end, the social impact end of the trust spectrum, the discussion mostly has relied on treating AI (esp. GenAI) as natural systems of sorts. That is, most of the efforts are placed in understanding of the AI models already in works (without much guardrails) and their implications often on specific use-cases driven by isolated examples. Many a discussions on both academic and policy forums, as well as public forums can be found trying to build incoherent understanding of issues arising from these deployed models and just as widely, making less-than-informed case for the benefits of these models. Most such discussions either end up being hyperbole or marginal. Both lack nuance, comprehensiveness and coherency.\footnote{This is made worse by a systemic failure of our social structures to value content-based discussions in favor of discussions measured on a \textit{popularity} scale whether it is via \textit{influencer} status on public domains or meaningless metrics such as \textit{citation counts} in academic and scientific circles. However, we do not delve into such issues since these are beyond the scope of this article.}

The effects and implications of AI on society are not a consequence of some abstract measures of trust but of the real-world outcomes that result from its use. Neither of the above set of efforts are anchored in any \textit{context} which goes to explain why they have encountered limited success in achieving the purported policy goals. Without context, \textit{trust} may be measured or quantified based on arbitrary metrics, but will be rendered meaningless when it comes to addressing the core issues that confront the society. This context is provided by specific utilization of the technology and  determines whether the effects are beneficial or deleterious to the society and stakeholders. In a range of other areas (both technological and policy), we have often defined this context via preferred outcomes. The goal of any policy making is to ensure that the outcomes of endeavors are aligned with agreed-upon societal values and priorities that improves the lives of the citizens. As the risk level increase in various context of application and adoption of AI across sectors, we need matching trust measures determined by the utilization in respective contexts. Hence, Fig.~\ref{fig1} also provides a proxy for how the importance of trust grows with adoption and that this exercise in building trust in technology should focus on the desired results for specific applications determined by relevant stakeholders.

\subsection{Functional components of the framework}

\begin{figure}
\centering
\includegraphics[width=425pt,height=225pt]{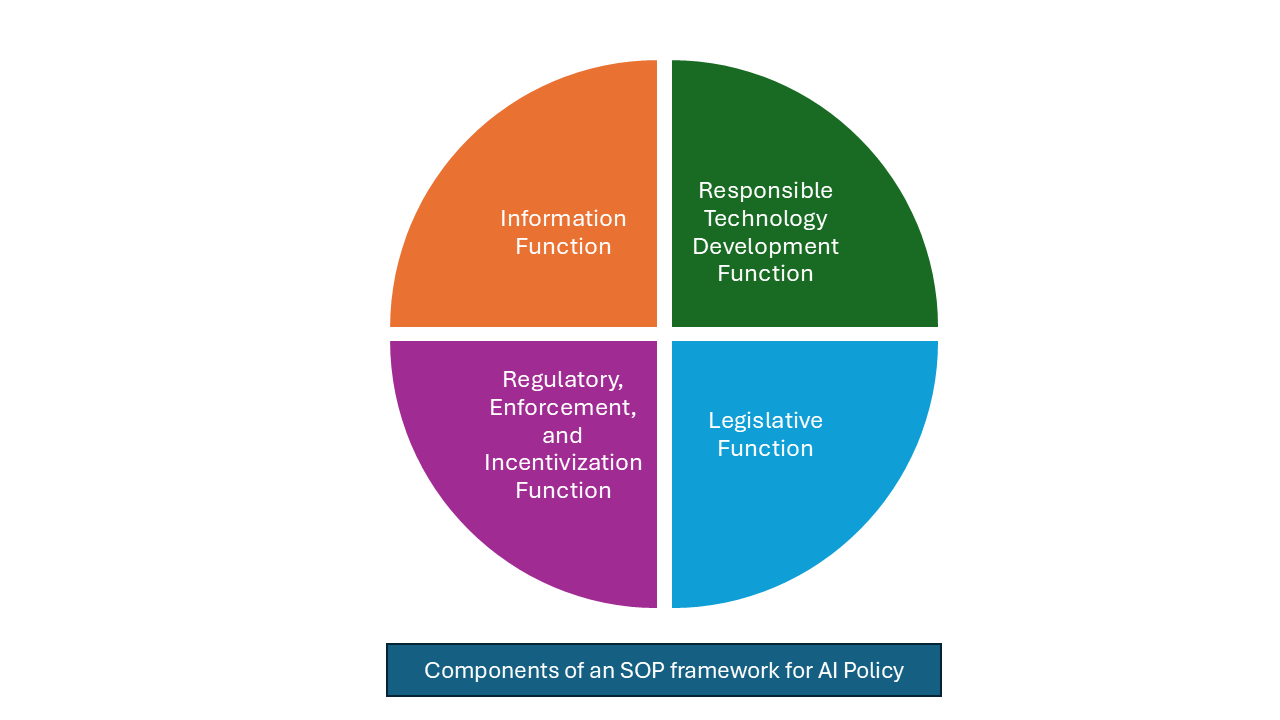}
\caption{Components of an SOP framework for AI Policy}
\label{fig-sop-components}
\end{figure}

An SOP framework consists of four core functions shown in Fig.~\ref{fig-sop-components}. It is an evolving framework - a \textit{live} policy regime keeping up with various technological and social developments, understanding of social and national priorities, effects and implications of both the developments and policies on society and the actions needed to expand and/or course-correct the policy and regulatory framework, all \textit{in the context of} consensus driven social priorities. Consequently, the first core function is an \textit{information function} that fulfills this need, providing an objective understanding and visibility to the policy efforts in an ongoing manner.

It is also important to have an evolving understanding and consensus allowing for responsible development of AI technologies along with the associated data-sensitivity, safety and risk-containment provisions balancing social good and also encouraging continuous innovation. This is the second core function - a \textit{responsible technology development} function that focuses on technology development aspects along with the associated guardrails, data-regulations, safety and risk avoidance at the core technology level aligning it with national, and social interests. 

The above two functions need to inform policymakers to craft reasonable and effective policies for the areas not covered by the existing policy framework, reinforce current legislation and address gaps on the policy and legislation front. Consequently, a \textit{legislative function} forms the third pillar of an SOP framework.

Finally, we need to make sure that AI-powered products and services are introduced in a responsible manner, and safeguard social interests, safety, privacy and other core domain-relevant concerns. The \textit{regulatory, enforcement and incentivization function} makes sure that various AI products are safe and beneficial when released and adopted across various areas of society and also create incentive structures to encourage meaningful developments.

These core functions and proposals for their implementation are detailed below.

\subsubsection{Information Function}

Any policy effort is hinged on accurate, objective, comprehensive, holistic information supported by real-world evidence. It is important that all the stakeholders and society in general participate in this critical discourse based on common unbiased set of facts and projections. Further, it is also important that the data and analyses that inform the policy go beyond one off concerns and assess and evaluate dependencies, future risks, opportunities as well as the implications of any policy proposals and actions on critical areas. There is no entity presently that serves to comprehensively inform the AI policy discourse. Consequently, not only are we in a continuous reactive mode to address isolated challenges and focusing on short term opportunities, we are also unable to see the broader view that provides insights into associated dependencies, and impacted areas that have been either neglected or overlooked. We lack a coherent and common understanding of the realistic and important opportunities or challenges.

Let us consider a couple of examples:
\begin{enumerate}
    \item The energy demands for AI development and maintenance efforts have been growing exponentially and are projected to continue to do so. There have been various proposals and efforts on building new energy-intensive data centers, along with respective energy-sources. These data centers have significantly high energy needs and operational resource costs~\cite{belanger2024,ren2023}. When considering these proposals for energy provisioning in light of arguments such as maintaining national competitive advantages or technological prowess, it is also important to understand their impact -- both current and future since the energy needs continue to grow -- on other societal priorities on sustainability, climate, localities with power centers, impact on other areas and businesses with energy needs and broader impact on the tax-payers~\cite{castelvecchi-2024}.\footnote{Many proposals on the table demand a public-private partnership to support the energy and other data-center needs. Further, there can be indirect cost impacts on the energy needs and energy consumption of businesses and consumers.} Such comprehensive understanding will not only allow for the policies to optimize and appropriately prioritize support for these needs but also balance them against the relative advantage to societies, and address impacted areas proactively.
    \item AI can have significant impact on the current and future labor markets. The main thrust across the business enterprises forming the bulk of value-generation is in productivity gains - that is automation resulting in reduction of the manual labor force. Having a realistic understanding and projected trajectory of labor market impact should allow for informed policy proposals not just on addressing short term labor market volatility but also for building strategies to address long-term impacts in time to minimize the labor market disruptions and displacement ensuring a socially smooth transition. It will also enable building relevant safety-nets for marginalized groups of stakeholders as a result of AI adoption.
\end{enumerate}

This information function can take many forms. We propose to establish \textit{A Congressional AI Office} -- an \textit{independent, strictly non-partisan, competence-based} congressional agency providing an \textit{objective} understanding of various aspects of AI landscape to both the congress and general public. It will consolidate works from various sources (e.g., US AI safety institute, NIST, and academic findings), collaborate and coordinate with other agencies and departments as well as perform independent research to understand, and envision AI developments, opportunities, risks and implications for various functions of US government and society.  It will publish these findings, papers, and recommendations on areas of national and social importance. The artifacts from this office will also inform national innovation prioritization by US government and coordinate with various agencies such as NSF and DARPA. It will also examine the relationship and interdependence of AI with other emerging and existing technologies as well as analyze existing policies and their effectiveness in achieving the original objectives. That is, this office will keep us \textit{honest} about the potential, risk, and policy effectiveness around AI. Further, it will also look ahead on the emerging risk landscape and leading indicators for concerns.\footnote{see, for instance,~\cite{koebler2024,speer2024} on the impact of GenAI on human language data;~\cite{ren2023} for sustainability related impacts on water usage. These issues are currently not on the policy radar.}

This entity will work with other agencies, policy makers, and departments to outline the state of AI landscape, coordinate and reconcile standards, make informed studies and policy recommendations, and potentially provide inputs in tracking of, and enforcement around, core AI technology developments. With the inputs from this agency, other respective departments and agencies can incorporate sector-specific AI policies, guidelines, and regulations. Further, it will also be a conduit for the much needed reconciliation across the federal and various state governments' efforts.

\subsubsection{Responsible Technology Development function}

There are various aspects of technology development itself that demand careful considerations from a policy perspective. The \textit{responsible technology development function} will assess and evaluate technology priorities in terms of development, opportunities, risks, standardization, privacy and security, national and social interests, impacts and dependencies. This will include the entire technology development pipeline -- that is, it will not focus just on the AI model development but will also encompass the associated domains including hardware, energy, resources, embedded systems, software as well as standards and guidelines for safe and responsible development, safeguarding, and data use. 

This function will hence coordinate and manage ongoing efforts across various branches of administration to ensure safe and responsible AI technology development along with ensuring the safety, privacy and safeguarding core issues of national and social interests. It will also be informed by other functions within the SOP framework in terms of both the priorities and downstream effects of the choices made by this function.

We propose to implement this function via a federal -\textit{US Data and AI Safety Agency} with regulatory and enforcement authority. It will focus on the technological aspects of AI life-cycle informed by the Information function and its own research and analyses efforts. It will also act as a central coordinator and consolidator of work done currently by various agencies and institutes such as the US AI Safety institute, NIST, and other departments (e.g., Department of Energy) and national labs, and also the private sector and academia to build a coherent strategy driving meaningful complementary efforts across the AI development life-cycle. Consequently, this function will provide an anchor for efforts such as guidelines on standardization (e.g.,~\cite{nist-ai-rmf-2023}), regulation and enforcement on responsible technology development as well as inputs to policy proposals around fundamental technology development areas (e.g., data privacy bill proposals, and AI safety proposals) that can inform congress and state legislatures for effective policy formulation.

\subsubsection{Legislative function}

The \textit{legislative function} naturally falls in the ambit of congress and policymakers. However, in addition to policy and legislative proposals on technology itself, this function will focus more importantly on the desired set of social outcomes and mitigating against risks to those outcomes as informed by the information function. Hence, it is likely that such legislative proposals, in service of ensuring beneficial outcomes, may go beyond AI and cover other technological and regulatory areas. For instance, while mis- and dis-information are certainly a risk, having disparate policies on deepfakes, or against spreading incorrect and unverified information, or election interference using mis- and dis-information, are piecemeal solutions. The desired social outcome should focus on the \textit{right to factual information} and a legislation to \textit{ensure information veracity, correctness and provenance} would be needed to bring coherence to this discussion and more importantly to make sure that the efforts can result in meaningful desired outcome. Such a proposal would then not only guard against the use of AI for deepfakes but will also cover other aspects such as the use of intentional misinformation, the role of algorithmic techniques in the dissemination and spread of such misinformation, and also the use of non-AI means to spread mis- or dis-information (for instance, spreading information without proper context, reporting incorrect and/or unverifiable statistics, unverifiable reports about healthcare, and so on). Such a need for factual information based discourse has already been proven critical not just in the context of free and fair elections but also pandemic awareness. More importantly, reliable information should be a fundamental right of the citizenry. Hence, policies to guarantee reliable, verifiable information needs to be a continuous effort - not just specific event-centered one - to guard against systemic risks (such as impact on social harmony, distrust in institutions, and social engineering). Similarly, while there are various efforts on guarding user data and associated privacy, they still fail to address the cumulative effects of such data collection on various societal functions that can range from limiting consumers' and citizens' options all the way to mass surveillance. Hence, it is important that the policies focus on the desired outcomes in an \textit{outside-in} manner than be technology-centered. The legislative function will need to deal with the AI challenges in their entirety in a significant departure from the current fragmented approach. There is another important aspect of making policies for the information ecosystem in that the citizens' $1^{st}$ amendment rights should be respected and protected at all costs. This has been one of the biggest continued challenges for the policymakers. An SOP framework has the advantage that it operates rather on the outcome space -- ensuring that the outcomes or consequences of the use of AI are in the societal interest without the need to employ constraints on the action space that may jeopardize citizens' rights and freedoms.

The legislative function's main operational role would be to bridge the gap between the existing policy and regulatory framework on social outcomes and novel challenges arising from AI and related areas. It needs to establish and/or reinforce legislation by identifying the areas that are either not covered and addressed by the current framework or where the current regulatory, legal, and enforcement structures need to be updated. Further, the legislation function isn't limited to guarding against risk. It also needs to actively build future readiness. Armed with broad stakeholder participation and a robust information function within the SOP AI framework, the legislative function would identify priority areas in social programs, research and development, national and foreign policies, humanitarian efforts, and diplomacy. The goal would be to not just be a reactive entity but to act as a catalyst towards desired social outcomes.

\subsubsection{Regulatory, Enforcement and Incentivization function}

As we focus on the desired outcomes at the legislative level, and ensure that the development of the core technology happens in a safe and responsible manner, the coherent policy framework makes it easier for various departments and agencies to propose effective, non-onerous and manageable regulatory frameworks, clear enforcement actions, as well as plans to incentivize and prioritize various programs in scientific research, industry support, innovation, and technology-integration and -adoption. Naturally, this \textit{regulatory, enforcement and inventivization function} is a distributed function shared across the bodies mandated with various important roles across the administrative spectrum.

Together with other constituents of the AI policy framework, it will help bring clarity for respective regulatory, enforcement and support agencies and departments such as the CFPB, SEC, FTC, NHSTA, FDA, DoD, DoT, Nuclear safety agency, NSF, and DARPA. Moreover, since the responsible technology function would have its respective regulatory and enforcement mandate at the technology level, these departments and agencies can focus on reinforcing and clarifying the AI integration within their respective mandates supported by objective data and analyses from the information function. Hence, these bodies won't have to worry about the technological aspects of AI individually. The cross-function information flow would allow for the technological concerns and application concerns to be directed to respective functions and inform relevant policy and regulatory efforts. An SOP framework will also empower these stakeholders to propose and implement policies within their respective mandates.

Below are a few examples of how this outcome-centric pivot within an SOP framework can be effective in various cases:

\begin{itemize}
    \item CFPB can have a say on avoiding unwanted outcomes and taking appropriate actions against behaviors such as unfair pricing, price gouging, customer burdened with junk fees, and customer discrimination. There is a significant risk potential using perfectly acceptable AI models but applied unethically by the industry. We have already witnessed some unfortunate examples such as~\cite{ross2023,viceuber2023,dubal2023,wray2024}. A technology-centric approach is currently pushing back these issues in some cases as AI algorithmic challenges which not only is sub-optimal but also helps maintain and potentially expand the risks of inappropriate employment of AI. Various agencies can implement appropriate framework for avoiding unethical or illegal behavior by actors incorporating any such efforts aided or abetted by AI in a relatively straight-forward manner.
    \item Adoption of AI in safety- , business- and mission-critical systems will still need to rely on the \textit{risk-adjusted outcomes} - for instance, use of AI in warfare will need to adhere to much more stringent guarantees and regulatory setup~\cite{humble-2024}. The use and risk tolerances of AI-driven products in different use-cases will naturally be subjective and should be decided by the criticality of the applications by bodies and experts in the respective domains to ensure safe and secure adoption.
    \item Integration of AI in consumer-safety related products and services will need to ensure the fair and ethical consumer outcomes  - for instance, NHTSA can evolve the existing framework for autonomous driving capabilities (e.g., for ADAS and autonomous vehicles) to incorporate AI capabilities in a coherent safety-based framework.
    \item Privacy focused efforts around consumer and proprietary data, whether publicly or privately available, can extend the guardrails \textit{beyond} their inappropriate use in AI model development and usage (using data during AI development can be different from that post-deployment which is not currently considered under data-protection frameworks; most efforts focus on avoiding use of user-data during model training).
    \item Guidelines, and if needed regulations, can be put in place for release, monitoring and risk-mitigation for AI products touching sensitive areas and sectors. A release plan similar to clinical drug trials can be thought of for sensitive domains that risk having large-scale and/or long-term impacts. Such a framework can provide graduated release of products much like pharmaceutical drugs. It should also enable monitoring the effects of technology post-release in a similar manner. And finally, this framework should mandate  \textit{recall} of products as and when necessary along with mitigation requirements. Depending upon the areas and the potential of risk identified with a specific domain and/or product categories, the respective frameworks can provide specific criteria for such planning and monitoring. This again can be similar to FDA working with the companies to establish end-points in clinical studies along with potential acceptability criteria for approval upon successfully meeting them in the trials.
    \item This SOP framework would also allow for regulating specific products, their specific features or potentially regulating them for specific uses (both allowing select use-cases and/or barring specific use-cases as relevant). As an example, AI algorithms that can generate images, text or videos on demand can be used for creative pursuits (e.g. design) as well as for damaging activities such as generating deepfakes. An outcomes based framework can address these by providing stringent guidelines and regulations on managing AI algorithms for various uses while also not limiting innovation.
    \item AI technology is also demonstrating implications on the open-source approach that has traditionally supported innovation and access. A balanced approach needs to ensure that we can maintain this accessibility but at the same time limit the negative consequences including:
    \begin{itemize}
         \item access to technology for nefarious and adversarial actors
         \item use of technology for purposes such as misinformation and disinformation by \textit{all} actors (for instance, there is potential and even evidence of use of AI for creating misinformation campaigns thereby threatening fair elections process in more than one countries)
         \item access to technology enabling potential for national security and corporate risks (e.g., by providing AI model details easily that may increase cyber-security risks)
         \item balancing global access with national competitive advantages
    \end{itemize}
    \item An SOP framework also allows to put other policies in perspective when trying to regulate AI. For instance, while the CHIPS act calls for limiting access to specific chips for adversarial state actors, it does not focus on the desired outcomes which is to limit the ability of these adversaries to develop technologies that can hurt democratic societies or become weapons of warfare. An outcome based framework also then forces us to think about access to capabilities that can bypass these actors' need for chips to begin with. For instance, one of the main uses of the compute infrastructure powered by specialized chips is to be able to train large AI models at scale. Hence, even if chips access is blocked, the policy does not place any limits on these actors' use of commercial cloud infrastructure~\cite{baptista-2024}. Consequently, the success in the efforts to limit AI advantage for the adversarial actors can't be achieved if they are able to bypass these restrictions. A fragmented approach hence proves to be ineffective and all the while places much more regulatory and compliance overhead on the industry and other stakeholders.
\end{itemize}

\section{Remarks}\label{sec:remarks}

The main objective of the SOP framework is to re-frame the discussion and execution around AI policy in terms of the associated outcomes for the society and align it with the values that we covet. For some time now, we have collectively been struggling to find an effective mechanism to ensure that the risks from AI are guarded against while leveraging and encouraging the developments for the betterment of society through meaningful research and applications. However, since the focus has been on the intricacies, working and managing the AI technological development itself, the policy efforts haven't helped us meaningfully move towards the stated policy goals. By emphasizing the context of application armed with objective evidence-based continuous information basis, an SOP framework can have many advantages over the current fragmented approach:

\begin{itemize}
    \item Having an outcome based framework would provide significant visibility and clarity to the technology and product developers as to how their products and offerings should behave in the \textit{respective} application context. This clarity will also extend to the industry moving for rapid product development and integration but with little guidance or guardrails. Moreover, it will also allow for integrating and reconciling technology- and domain-focused efforts, ensuring their synergy, and avoiding conflicts.
    \item The technology community will be able to better focus its efforts in building guardrails around the AI capabilities in respective contexts. For instance, an AI functionality used in recommending shopping products will have different risk tolerance compared to the same functionality used for safety-critical applications like healthcare. Naturally, the product guardrails for and beyond the AI models should differ for the two use-cases.
    \item Downstream issues such as bias, user-discrimination, information distortion, and consumer discrimination arising from AI-driven products and services will also be addressed at the respective offering levels since the application context drives how the AI models should be employed and how the products should behave. For instance, it is expected that an AI model doesn't discriminate among equally qualified candidates based on protected characteristics (gender, religion, ethnicity, etc.) in the context of hiring or providing insurance. However, an AI model deployed specifically to assess biased outcomes of policy or product may need to explicitly consider these characteristics to provide meaningful information (e.g., a model used to assess if there are systematic effects on specific protected class). Hence, generic guardrails placed at the model-level themselves will be ineffective (and often impractical from a technology perspective).
    \item Domains such as healthcare and transportation have also been struggling to address concerns arising from the introduction of AI and have focused on what should or shouldn't be allowed at the technology level. However, the manner in which the technological advancements are achieved leave little room to incorporate these concerns in any explicit manner, and even when done, comes at significant performance cost. An outcome based approach will allow for addressing these concerns at the product level. For instance, autonomous driving utilizes AI models in various tasks. It may be too difficult, if not impossible, to regulate these models at task-level granularity. However, what matters is the safety profile of an autonomous vehicle in real-world setting. Hence, rigorous requirements on validation and verification of autonomous vehicles, their safety and predictability are needed much as they are for non-autonomous vehicles. It would be much more effective to expand and strengthen the qualification requirements for vehicles. Further, the liability concerns can be dealt with independent of the AI issues. When looked at from an outcome lens, these can be addressed at the intersection of hardware-,and software-providers and the owner/driver behavior as relevant. In fact, the DoT is already moving in this direction gradually.\footnote{We do not delve into legal aspects of specific application areas since the focus of this article is to suggest a broad policy framework that would allow for specific legal frameworks to be developed or reinforced in the AI context.}
    \item Areas of more systemic and long-term significance such as education, legal and even finance can have a wider participation from core stakeholders in understanding and promoting the desired outcomes rather than blind adoption of technology followed by a reactive approach to managing the consequences.
    \item Broader concerns around society's stability and upholding and strengthening its core value systems can be viewed holistically, unlike presently. These include areas in information provenance and reliability, institutional integrity, trust in the institutions, and social harmony among others.
    \item National and corporate security and competitiveness related focus areas can be dealt with more efficiently and effectively.
    \item Having clarity on the current state of AI, and tracking its evolution along with other concurrent developments, allows for a significantly robust future readiness strategy.
    \item \textit{Reconciling Federal and State efforts}: While there have been various efforts both at the federal and state levels to build frameworks focusing on various aspects of AI and related data technologies, these are aimed at addressing parts of the challenge such as data privacy~\cite{cpra2024}, initiatives on AI safety~\footnote{See, for instance, National AI Advisory Committee (NAIAC), US AI Safety Institute, and \url{ai.gov}} and standardization~\cite{nist-ai-rmf-2023,nist-ai-rmf-playbook2023,isoiec420012023}. A coherent SOP framework will help reconcile these efforts and also thereby avoid duplication and save resources -- both taxpayer funds, and personnel. AI doesn't necessarily operate in accordance with geographical (e.g., state) boundaries and hence it is important and imperative to have a national consensus on the policies with room to adapt for specific localized areas of concerns as needed.
    \item \textit{Building international cooperation and understanding on AI:} As AI gets implicated in an increasing number of domains including highly sensitive areas such as warfare~\cite{humble-2024}, it also touches upon safety and national security domains with potential serious concerns around their use. While we are not yet at a stage where AI would be able to automatically escalate conflicts, there are many areas that require immediate attention and the need to start building international consensus on fair use of AI on areas with geopolitical implications including defense, weaponry, cyber-security, and space applications and deployments. An evolving international treaty framework can allow us to build guardrails around intentional or inadvertent misuse of AI to guarantee global stability and also develop appropriate rules of engagement. This will also have implications for maintaining and strengthening democratic structure, institutions, and societies. An SOP framework provides a strong basis for an ongoing international collaboration effort.
    \item \textit{Reconciling Industry efforts:} Just as there are government efforts to address the risks of AI and democratize the benefits, various industry efforts are underway to address these areas too. Some such efforts have been made in participation with the government as mentioned above. Additional efforts are being driven either as initiatives of industry groups or various technology companies. While the former takes the form of potential agreement on specific outcomes (e.g.,~\cite{c2pa2024,isoiec420012023}), the latter range from building guardrails on existing technologies (e.g., efforts from large technology companies around unintended or unsafe use of AI) to trying to build novel products with some specific safety by design (e.g.~\cite{constitutionalai2023}). As the scope and scale of AI technology expands, esp. with new and rapid developments in generative AI offerings and their subsequent adoption for various applications, it is important that there is a collective understanding on the minimal safety standards for AI products. It is certainly a non-trivial task, but it is important to address these challenges as soon as possible. Further, the industry and government efforts should cross-inform each other to make sure that they do not diverge. An SOP framework would add a strong information basis and help build this reconciliation across various efforts.
    \item \textit{Participation from key non-AI stakeholders and society at large:} So far, \textit{the participation and representation of the wider societal stakeholders has been limited} when it comes to shaping AI policy. There is certainly a significant value in having AI and industry representation as these stakeholders have significant power and responsibility to effect change in a productive manner. However, there is a much larger constituency that is and will increasingly be on the receiving end of the AI's outcomes, impacts, and consequences of the policy decisions. This broad constituency so far has found limited representation in the discussions, and both the industry and the government should enact mechanisms to enable their participation in the decision-making process. For instance, these representations can come from social scientists, economists, ethicists, labor representatives, and general public whose inputs should inform the comprehensive policy framework, esp. around the desired outcomes. A candid and transparent approach is also in the interest of the AI industry and proponents to build public trust, and hence buy-in, for the adoption of AI.\footnote{Some industries have already started realizing the value and need of building this trust for adoption and acceptance of AI offerings~\cite{goldman-2024}.} An SOP framework provides clear mechanisms for such participation from all the stakeholders. 
\end{itemize}

\section{Call for Action}\label{sec:conclusion}

AI technology has scaled rapidly in the past years bringing in significant interest from various spectra of society searching for solutions to important problems. At the same time, there is significant perceived opportunity in the industry for monetizing the technology in various ways. This has resulted in a race to realize these opportunities. Unfortunately however, the discussion, policy, and operationalization of these efforts in socially beneficial ways has taken a backseat. The policy makers have been taken by surprise with the pace of technology development, some promising breakthroughs, and the already evident risks and consequences of rapid adoption of AI. In the absence of any robust guidance, there is a clear lack of checks and balances that can ensure safety, security and the right direction of these developments and adoption of technology. The discussion on building these checks and balances has been suffering due to various factors: the hyperbole resulting from unverified claims of AI promise, the dire warnings around a hypothetical future risk of AGI and ASI distracting from the risks and challenges in the present, the narrow focus on specific AI outcomes due to a highly divided information-challenged social discourse, the lack (even absence) of nuance in the debate on the current state and projected trajectory of AI, lack of broader participation from non-AI stakeholders who are at the risk of being most impacted, non-objective and vested interests affecting or even driving proposals on AI policy, and disproportional weighting of opinions from "aura"-based non-experts (aka influencers). 

These factors have naturally hampered the much needed holistic discussion and an evidence-based realistic view of the issues. Also, the discussions around safe and meaningful adoption of technology have been stuck in a narrow technology focused narrative. Consequently, the policy efforts have been fragmented and in many cases decoupled from the intended social outcomes, the real policy goals. 

This paper has highlighted the limitations of the current technology-centered view on AI policy and surfaced the gaps that make this relatively myopic policy effort suboptimal at best. In fact, there are significant risks in continuing with the current approach as we have detailed above. We also make a \textit{case to pivot to a society-centered approach} on AI policy and propose a Social Outcomes and Priorities centered (SOP) framework. We presented the core components of such a framework along with initial proposals to implement them.

\textit{We have a limited window to effect meaningful AI policy framework} given how deep and wide impact the technology continues to demonstrate. We \textit{can not afford to be distracted} by narratives at the extremes -- that of making cases for unhinged, unregulated development and adoption or that of a hypothetical existential risk due to a super-intelligent version of AI which is imminent (\textit{it is not; not even close}). While we certainly need to be ready for future, we urgently need an effective meaningful AI policy framework to address the ongoing challenges and risk. Further we need to do so in a manner that is innovation-friendly, complements the scientific and technological progress, is minimally onerous and invasive, and is pro-active, adaptive and forward-looking. \textit{A pivot to a society-centered effort is absolutely critical, and urgent} for a meaningful, effective and efficient AI policy. \textit{Our inability to effectuate meaningful policy will be adversely consequential to all life on the planet.}

We hope that the current proposal becomes a basis to initiate a wider, and more consequential effort to achieve such meaningful AI policy. Let us bring back much needed thoughtfulness, reason, nuance, objectivity, transparency, intellectual and scientific rigor, and social responsibility to this very important discourse of our times.

%\backmatter
%\bmsection*{Author contributions}

%This is an author contribution text. This is an author contribution text. This is an author contribution text. This is an author contribution text. This is an author contribution text.

\bmsection*{Acknowledgments}

The author would like to thank Nathalie Japkowicz and Unmesh Kurup for helpful feedback on the article. The author would also like to acknowledge the broader AI, policy making, academic and industrial leadership communities for helpful discussions along the way.

Manuscript prepared using adapted Wiley's New Journal Design (NJD) LaTeX template.

%\bmsection*{Financial disclosure}

%None reported.

%\bmsection*{Conflict of interest}

%The author declares no potential conflict of interests.

\bibliography{Outcome-oriented-AI_Policy}

%\bmsection*{Supporting information}

%Additional supporting information may be found in the
%online version of the article at the publisher’s website.

%\nocite{*}% Show all bib entries - both cited and uncited; comment this line to view only cited bib entries;

\bmsection*{Author Biography}

\begin{biography}
{\includegraphics[width=76pt,height=80pt]{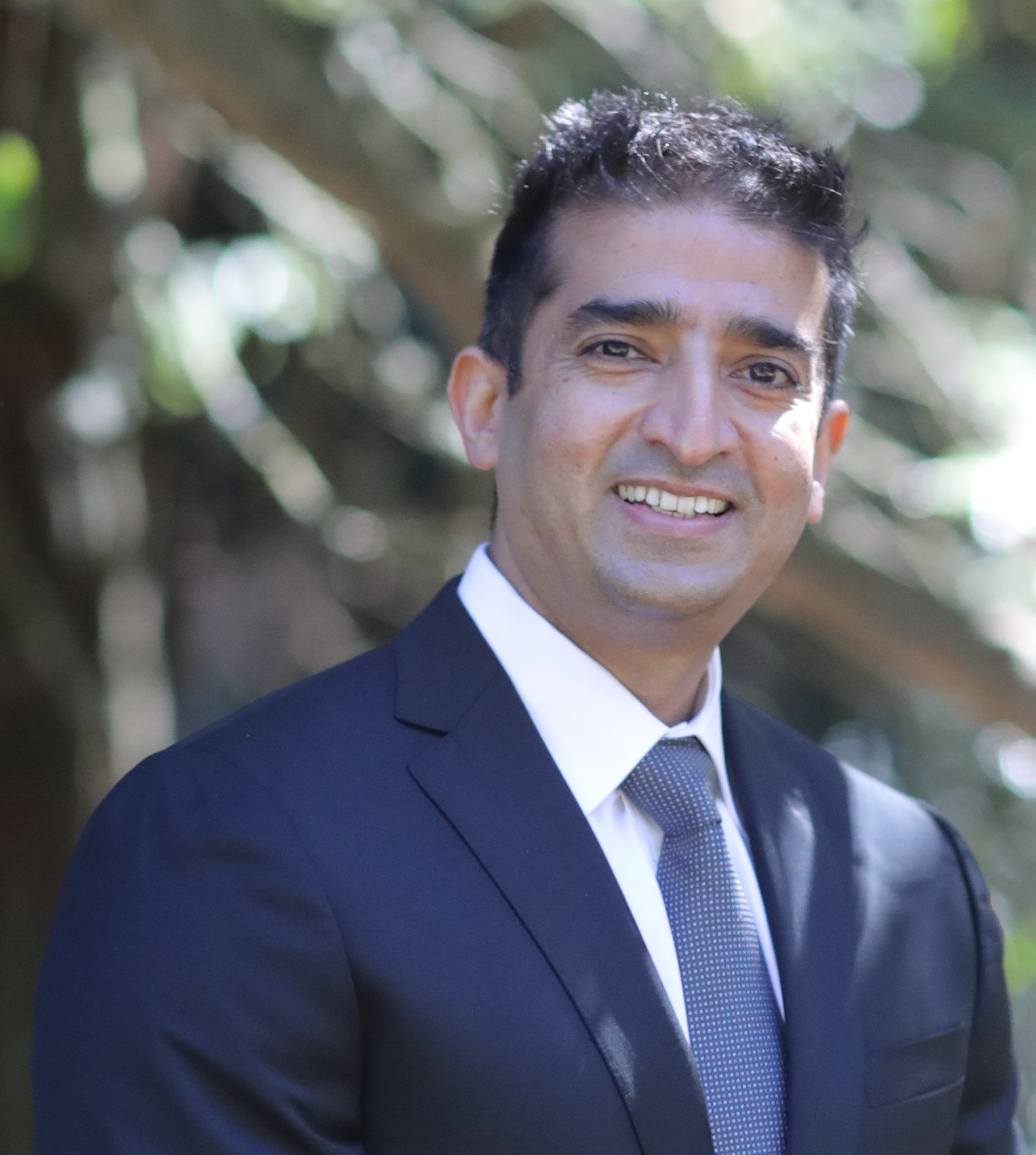}}
{
{\textbf{Mohak Shah} is an AI and technology executive with extensive experience in bringing data- and AI-products to market. He has held several senior leadership roles in large enterprises and startups driving both large-scale AI transformation initiatives, and zero-to-one product journeys. He is the founder and Managing Director of Praescivi Advisors, a strategic AI advisory practice, and regularly consults in strategic AI areas and AI transformation initiatives. As a research scientist, Mohak has published extensively in theoretical and applied machine learning areas.}}
\end{biography}

\end{document}